\documentclass[acmsmall]{acmart}

\usepackage{graphicx}
\usepackage[caption=false,font=footnotesize]{subfig}
\usepackage{multirow}

\usepackage{algorithm}
\usepackage{algpseudocode}
\usepackage{booktabs} 
\usepackage{tcolorbox}
\tcbuselibrary{skins}

\usepackage{xspace}
\def \projectName {UNICS\xspace}

\newtcolorbox{mybox}{
colback=gray!20,
colframe=gray!40,
boxrule=0.5pt,
boxsep=2pt,
arc=4pt,
left=3pt,
right=3pt,
top=3pt,
bottom=3pt
}
\usepackage{enumitem}
\usepackage{url}

\newtcolorbox{mybasecolorbox}[1][]{%
  colframe=gray!75,
  coltitle=white,
  fonttitle=\bfseries,
  width=(\linewidth),
  title=#1}

\newenvironment{mytitlebox}[1][]{%
  \centering
  \mybasecolorbox[#1]
}{%
  \endmybasecolorbox
}

\AtBeginDocument{%
}

\setcopyright{cc}
\setcctype{by}
\acmDOI{10.1145/3797067}
\acmYear{2026}
\acmJournal{PACMSE}
\acmVolume{3}
\acmNumber{FSE}
\acmArticle{FSE005}
\acmMonth{7}
\acmSubmissionID{fse26maina-p88-p}
\received{2025-09-12}
\received[accepted]{2025-12-22}

\begin{document}

\title{UNICS: Multilingual Code Search via Unified Pseudocode and Contrastive Transfer Learning}

\author{Ye Fan}
\orcid{0000-0003-3584-9920}
\affiliation{%
  \department{State Key Laboratory for Novel Software Technology}
  \department{Department of Computer Science}
  \institution{Nanjing University}
  \city{Nanjing}
  \country{China}
}
\email{yefan@smail.nju.edu.cn}

\author{Jidong Ge}
\orcid{0000-0003-1773-0942}
\authornote{Corresponding author.}
\affiliation{%
  \department{State Key Laboratory for Novel Software Technology}
  \department{Department of Computer Science}
  \institution{Nanjing University}
  \city{Nanjing}
  \country{China}
}
\email{gjd@nju.edu.cn}

\author{Chuanyi Li}
\orcid{0000-0001-9270-5072}
\affiliation{%
  \department{State Key Laboratory for Novel Software Technology}
  \department{Department of Computer Science}
  \institution{Nanjing University}
  \city{Nanjing}
  \country{China}
}
\email{lcy@nju.edu.cn}

\author{LiGuo Huang}
\orcid{0000-0001-7790-0195}
\affiliation{%
  \department{Department of Computer Science}
  \institution{Southern Methodist University}
  \city{Dallas}
  \country{USA}
}
\email{lghuang@lyle.smu.edu}

\author{Bin Luo}
\orcid{0009-0001-1102-9584}
\affiliation{%
  \department{State Key Laboratory for Novel Software Technology}
  \department{Department of Computer Science}
  \institution{Nanjing University}
  \city{Nanjing}
  \country{China}
}
\email{luobin@nju.edu.cn}

\renewcommand{\shortauthors}{Ye Fan, Jidong Ge, Chuanyi Li, Liguo Huang, and Bin Luo}

\begin{abstract}
    While pre-trained models have achieved remarkable success in code search, their multilingual capabilities remain a major hurdle, plagued by data imbalance, cross-lingual semantic interference, and the loss of critical information from existing unified representations like Abstract Syntax Trees (ASTs) or Intermediate Representations (IRs). Furthermore, conventional contrastive learning strategies often rely on simplistic hard negative sampling while overlooking the potential of mining hard positives to learn code's intrinsic semantic invariance. To address these challenges, we introduce \projectName, a framework for multilingual code search built on a two-stage training strategy. In the first stage, \projectName is pre-trained on a novel dataset we constructed, which uses pseudo-code as a unified representation to learn a cross-lingual, algorithm-level logic that preserves full semantic fidelity. The second stage employs a multi-task transfer learning strategy that adapts this general knowledge to specific languages by decomposing code into semantic slices (e.g., API calls, function bodies) and incorporating tasks for hard positive mining and cross-lingual dynamic hard negative sampling. Experimental results demonstrate that \projectName achieves state-of-the-art performance across multiple multilingual and cross-lingual benchmarks, showcasing superior generalization and performance balance, especially in zero-shot transfer tasks to low-resource languages.

\end{abstract}




\maketitle

\section{Introduction}
Code Search, which aims to retrieve functionally relevant code snippets from vast codebases based on natural language queries, is a key technology for enhancing developer productivity and promoting code reuse~\cite{BrandtGLDK09,CodeHow}. In recent years, the rise of deep pre-trained models has significantly advanced code search technology, with performance far surpassing traditional text-matching methods~\cite{intro:ir3,intro:ir1,intro:ir2,intro:ir4,intro:ir5}. In downstream tasks such as API recommendation, code completion, and bug fixing, using retrieval models to enhance contextual information has become the mainstream choice, outperforming conventional solutions~\cite{intro:ir7,intro:ir8,intro:ir9,intro:ir10,intro:ir11,intro:ir12}. Particularly in the wave of agent-based automated software development, leveraging Retrieval-Augmented Generation (RAG) to dynamically organize context has become a standard configuration and critical infrastructure for assisting code generation~\cite{CPM7,ijcai2022p775}. However, while existing models perform exceptionally well in specific languages, their multi-lingual code retrieval capabilities face severe challenges in real-world development scenarios~\cite{intro:Linstead09Sourcer,intro:deep1,intro:deep2,intro:deep3,intro:deep4,intro:deep5}.

Multi-lingual retrieval requires a model to accurately identify relevant code within a mixed-language codebase. In large code communities like GitHub and Stack Overflow, as well as in enterprise-level internal repositories, code resources often coexist in multiple languages. Neglecting non-mainstream languages can lead to a sharp decline in retrieval quality, preventing users from obtaining desired results.

Furthermore, in smart IDE environments, both code completion tools and development agents need the ability to search within a user's local repository. This means the model must handle complex information, including a mix of programming languages (e.g., Python, JavaScript) and configuration files (e.g., JSON, YAML, HTML), and even semi-structured text like README documents, developer tutorials, or issue logs.

Therefore, the multi-lingual scenario is not only a touchstone for evaluating a model's generalization ability but also a core obstacle that must be overcome for practical application. To achieve robust multi-lingual retrieval, a model must learn to distinguish and understand diverse programming syntaxes and paradigms during the pre-training phase. However, current research faces several major bottlenecks.

The primary challenge in achieving powerful multi-lingual retrieval capabilities lies in the scarcity and imbalance of training data. The high cost of annotating high-quality "code-query" pairs means that existing mainstream pre-training datasets (e.g., CodeSearchNet, CoSQA) only cover a few popular programming languages like Python and Java. This data bias causes existing models to exhibit significant performance imbalances in multi-lingual environments, making it difficult to effectively generalize their capabilities to low-resource or niche programming languages not present in the training data.

To mitigate the data scarcity problem, researchers often employ transfer learning strategies, pre-training a model on data from popular languages and then fine-tuning it on a target language (e.g., using code translation datasets or niche language datasets). However, this seemingly straightforward solution is often plagued by data noise and semantic interference, and can even lead to catastrophic forgetting, which degrades the model's performance on the source languages.

\textbf{Data Noise:} The methods used to construct cross-lingual datasets introduce a significant amount of noise. For example, many code translation datasets rely on automated tools that use lexical or syntactic similarity to match functionally equivalent snippets in different languages. These heuristic methods are not rigorous and introduce many false positives where the code snippets are not functionally matched.

\textbf{Semantic Interference:} Different programming languages have vast differences in syntax, naming conventions, and programming paradigms, and direct transfer can cause semantic interference. For instance, the keyword `let` defines a mutable, block-scoped variable in JavaScript, whereas in Rust and Swift, it denotes an immutable binding. Similarly, the `static` keyword has different scope and lifecycle implications in C++ and Java. When a model learns patterns from one language, it can easily conflict with its knowledge of another.

This conflict has prompted researchers to ask: can we find a Unified Representation for code that bypasses the surface-level differences between languages to achieve more efficient and lossless knowledge transfer?

Some research has attempted to use graph-based intermediate representations (IR), such as Abstract Syntax Trees (ASTs) or Control Flow Graphs (CFGs), to unify code from different programming languages~\cite{ma2023mulcs}. The intention is to enable the model to explicitly learn the syntactic structure and execution logic of the code, such as conditional branches, loops, and return paths within a function. This does provide the model with structural insights that a pure text representation lacks.

However, for tasks like code search that heavily rely on semantic understanding, this over-dependence on structure can be counterproductive. First, abstracting code into a graph structure inevitably leads to the loss of critical textual information. For example, indentation, which is crucial in Python, code comments, and the carefully chosen variable names by developers are all simplified or discarded in an AST. Second, API call sequences and specific keywords (like `async/await`) are often direct indicators of a code's core functionality, but in a graph representation, they may be relegated to mere node labels, diminishing their rich semantic weight.

Finally, the training paradigms of existing methods, especially those based on contrastive learning, also have significant limitations.

First, to teach the model to distinguish subtle differences between code snippets, many studies rely on training with hard negatives. However, their methods for generating these hard negatives are often too simplistic, such as random token replacement or simple in-batch sampling. This static or random negative sampling strategy overlooks a critical dynamic issue: the similarity of code representations continuously changes as the model trains. A negative example considered "easy" at the beginning of training may become difficult to distinguish from a positive example later on, and vice versa. A fixed sampling strategy cannot dynamically adapt to the model's evolution, often leading to overfitting on simple examples while failing to learn sufficiently from genuinely difficult ones.

Second, existing methods overemphasize learning "dissimilarity" from negative samples while neglecting to learn "invariance" from positive samples. In addition to the original positive pairs, we can generate functionally equivalent but representationally different hard positives through code transformations and slicing. Introducing these samples forces the model to learn more robust and abstract semantic representations, thereby effectively mitigating overfitting.

Finally, most contrastive learning methods are confined to a single-language setting, failing to fully leverage cross-lingual data. Functionally similar (positive examples) or functionally approximate (hard negative examples) code snippets exist across different programming languages. Ignoring these cross-lingual contrastive signals severely limits the model's multi-lingual understanding and generalization ability and wastes the valuable cross-lingual supervisory signals embedded in massive codebases.

In summary, no existing work has addressed the interference issues arising from multi-lingual settings, the information loss caused by unified IR representations, or the problems stemming from neglecting positive samples during training. Based on these observations, we propose \textbf{\projectName}, a transfer learning framework for multi-lingual code retrieval based on a unified representation. \projectName consists of two training stages: In the first stage, we construct a dataset based on a pseudocode-like unified representation of code. We use contrastive learning to pre-train the model, loading the knowledge from this unified representation. In the second stage, we employ a multi-lingual transfer learning approach. We use a code slicing method to split data from different languages into components with varying semantics. Through multiple tasks—including a slice-type prediction task, hard positive contrastive learning, and hard negative contrastive learning—we transfer the pre-trained knowledge from the first stage to different languages.

We compare \projectName with several state-of-the-art models and find that \projectName achieves SOTA performance on multi-lingual retrieval tasks. We observe that \projectName can be effectively transferred to niche language retrieval and mixed-language retrieval tasks with minimal transfer loss. Empirical studies show that \projectName exhibits far superior generalization capabilities across programming language scenarios with varying granularities and syntactic types compared to existing models.

In summary, we make the following contributions:
\begin{itemize}
    \item We propose the \textbf{\projectName} training framework. Through our designed lossless unified code representation and a multi-task, multi-lingual contrastive learning approach, we have developed a state-of-the-art (SOTA) multilingual code search model.
    \item We have constructed a dataset that includes a detailed design for the unified code representation. We have also innovatively designed multiple hard positive generation methods, multi-lingual pre-training tasks, and a dynamic hard negative learning method.
    \item Our experimental results demonstrate that our model achieves leading results in multi-lingual search and search tasks across various niche programming languages, with lower transfer loss.
\end{itemize}

The remainder of this paper is organized as follows.  We present the relevant work in Section~\ref{sec:rel}. Section~\ref{sec:method} overviews our proposed approach. The experimental setup and results are then described in Sections~\ref{sec:experimental_setup} and~\ref{sec:experiments} respectively.We discuss threats to validity in Section ~\ref{sec:threats} and conclude in Section~\ref{sec:conclusion}.

\section{Related Work}
\label{sec:rel}
\subsection{Code Pretrained Model}
In the field of code intelligence, it has become a mainstream paradigm to pre-train models on large-scale code corpora and then fine-tune them for downstream tasks, such as code retrieval. The core contribution of such research lies in designing diverse pre-training tasks aimed at learning unified and generalizable knowledge representations from code~\cite{ijcai2022p775}.

One line of work focuses on utilizing the structural information of code. For example, models like UnixCoder~\cite{UniXcoder2022}, SynCoBERT~\cite{SyncoBert2021}, and GraphCodeBERT~\cite{GraphCodeBERT2021} employ Abstract Syntax Trees (ASTs) to capture the syntactic structure of code. However, due to significant structural differences in ASTs across various programming languages, this approach struggles to learn universal cross-lingual structural knowledge. Its role is mainly limited to enhancing the model's understanding of the code structure of a specific language.

Another category of methods explores Intermediate Representation (IR). For instance, some models convert code from different languages into a unified IR and learn the internal structural labels and dependencies of the code using graph neural networks (such as GGNN)~\cite{ma2023mulcs,huang2023towards}. Although IR, to some extent, ignores the syntactic details of a language, experiments have shown that methods incorporating such structural information still achieve performance improvements over traditional approaches, which confirms the importance of structural information in code representation learning~\cite{DBLP:conf/wcre/WangLM0J20,DBLP:conf/icsm/SvajlenkoIKRM14}.

Furthermore, to learn multi-granularity semantic information from code, some models segment code into units of different granularities, such as lines, code blocks, and functions, and aggregates features through mean pooling~\cite{huang2023towards}. However, the performance improvement of this method is limited. We argue that a powerful pre-trained model should be able to adaptively learn semantic features at different levels during the training process, rather than relying on explicit manual segmentation~\cite{DBLP:conf/iwpc/ShuaiX0Y0L20,DBLP:conf/kbse/WanSSXZ0Y19}.

Inspired by the aforementioned work, this study aims to find a unified code representation that is more general-purpose and has less information loss. We propose using pseudocode as a normalized representation of code. Compared to AST or IR, this method preserves the semantics of the original code to the greatest extent while maintaining language-agnosticism. Additionally, we have designed a semantic-based segmentation strategy instead of a simple physical size-based one. This strategy can decompose code into units with different semantic functions, thereby achieving more refined feature learning~\cite{COCO2022old,COCO2022new,CodeBERT2020,CodeGPT2021,CodeT52021,ContraCode2021,deepcs}.

\subsection{Multilingual Code Retrieval}
In the domain of multilingual code retrieval, existing research primarily extends single-language models by introducing specific pre-training tasks to enhance the model's cross-lingual understanding and alignment capabilities.

Some studies focus on mitigating the data imbalance problem. For example, some work has noted significant performance disparities of models across different languages and, for this reason, introduced a language label prediction task to enhance the model's ability to discriminate language features, while also adjusting the sampling ratio of each language during data loading~\cite{li2024consider}. However, this strategy offers limited benefits for improving the model's generalization ability to unseen languages and may violate the fundamental assumption of independent and identically distributed (i.i.d.) training data, thereby harming the model's overall generalization performance~\cite{zhu2022xlcost}.

A portion of the work is dedicated to optimizing the contrastive learning framework. For instance, LamCODE~\cite{huang2023towards} uses functionally matching code pairs as positive samples for contrastive learning and combines this with a random masking strategy for training, but this method risks introducing noisy samples. Contriever~\cite{izacard2021unsupervised}, on the other hand, employs in-batch and cross-batch negative sampling strategies. We believe that such static negative sampling methods fail to fully utilize all available negative samples, because as the model parameters are iteratively updated, the distribution of hard negatives also changes dynamically. In view of this, this paper proposes a dynamically updated cross-lingual negative sample mining mechanism, which has achieved better training results~\cite{rerank21robust,relate:contra2}.

Other research has attempted to leverage cross-lingual code translation datasets to improve the model's cross-lingual capabilities. For example, CodeRetriever constructs a parallel code corpus by retrieving similar documents and function names~\cite{CodeRetriever2022}. However, this construction method lacks strict alignment guarantees, is prone to introducing noise, and its applicability is limited to specific programming languages and scenarios, making it difficult to generalize to low-resource languages~\cite{dahal2022scotch}. Some approaches utilize translation data of human languages for learning~\cite{UniXcoder2022}, but we believe that this type of method offers limited help for the multilingual code retrieval task. The fundamental reason is that the differences between programming languages are mainly reflected in syntactic structure, keywords, and API calls, rather than the diversity of natural languages—most programming languages are still based on English lexicographically~\cite{yan2023codetransocean}.

In comparison, the method proposed in this study has two major advantages: first, the pseudocode representation we adopt is a nearly lossless unified paradigm, with information fidelity far exceeding that of code translation datasets; second, we have designed targeted pre-training tasks to specifically learn the differences in syntactic structure and API usage across different languages, thereby more directly and effectively enhancing the model's cross-lingual code understanding and retrieval capabilities.

\begin{figure}[t]
  \centering
  \includegraphics[width=.8\linewidth]{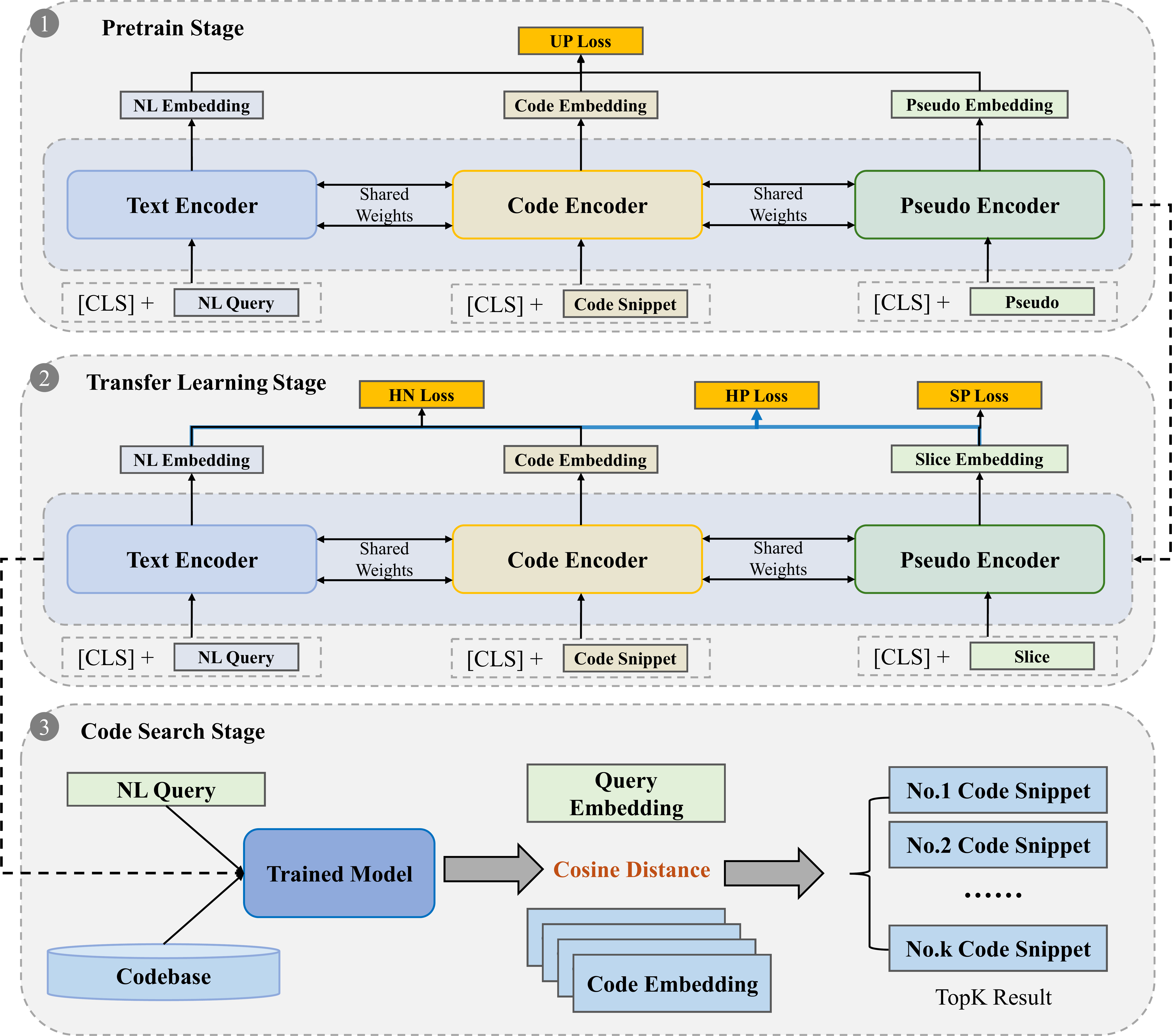} 
  \caption{The overall workflow of \projectName, which includes a pre-training stage, a transfer learning stage, and a final code search (inference) stage.}
  \label{fig:framework}
\end{figure}

\section{Method}
\label{sec:method}

In this section, we introduce the methodology of \projectName for learning the differences between various programming languages. As illustrated in Figure~\ref{fig:framework}, the \projectName workflow begins with dataset construction and is centered around two core training stages: Pretraining and Transfer Learning, followed by a final Code Search (inference) stage. We will detail each stage in the following sections.

\subsection{Pretraining Stage}

\subsubsection{Definition and Design Principles of Universal Code}
\label{sec:unidef}

\textbf{Universal Code} is a standardized representation for expressing algorithmic logic. It describes each step of an algorithm in a manner close to natural language and mathematical formulas, aiming to abstract away the syntactic details of specific programming languages and the complexities of machine implementation, while fully preserving the essential algorithmic logic. This characteristic makes it an ideal bridge between human thought and program implementation, often used in algorithm teaching and software development documentation.

By abstracting code from different programming languages into a unified pseudocode, we can construct a \textit{Unified Semantic Carrier}. This carrier effectively eliminates the surface-level syntactic differences among various programming languages, achieving deep cross-lingual semantic alignment while ensuring the process is lossless.

To ensure that the generated pseudocode meets high-quality standards, we have established the following three core design principles:

\begin{itemize}[leftmargin=*]
    \item \textbf{Lossless:} The pseudocode must completely preserve all core semantic elements of the original code. To achieve this, we not only require the use of keywords that can express rich semantics but also stipulate that the model must clearly describe key implementation logic in comments, ensuring every detail of the algorithm is faithfully reflected.
    \item \textbf{Consistent:} For code snippets that have the same functionality but different implementations (e.g., loops or API calls in different languages), the generated pseudocode must be as consistent as possible. To this end, we require the model to eliminate language-specific information, such as converting a specific API call (e.g., \texttt{requests.get}) into a natural language description of its function (e.g., \textit{make an HTTP GET request}), and expanding common abbreviations in the code to make its expression closer to general natural language.
    \item \textbf{Faithful:} The generation of pseudocode must strictly adhere to the semantic boundaries of the original code, without introducing any external information or making out-of-scope inferences. We require the model to ensure through repeated verification that every statement in the pseudocode has a clear one-to-one correspondence with a logic block in the original code.
\end{itemize}

\begin{figure}[h!]
\centering
\includegraphics[width=.7\linewidth]{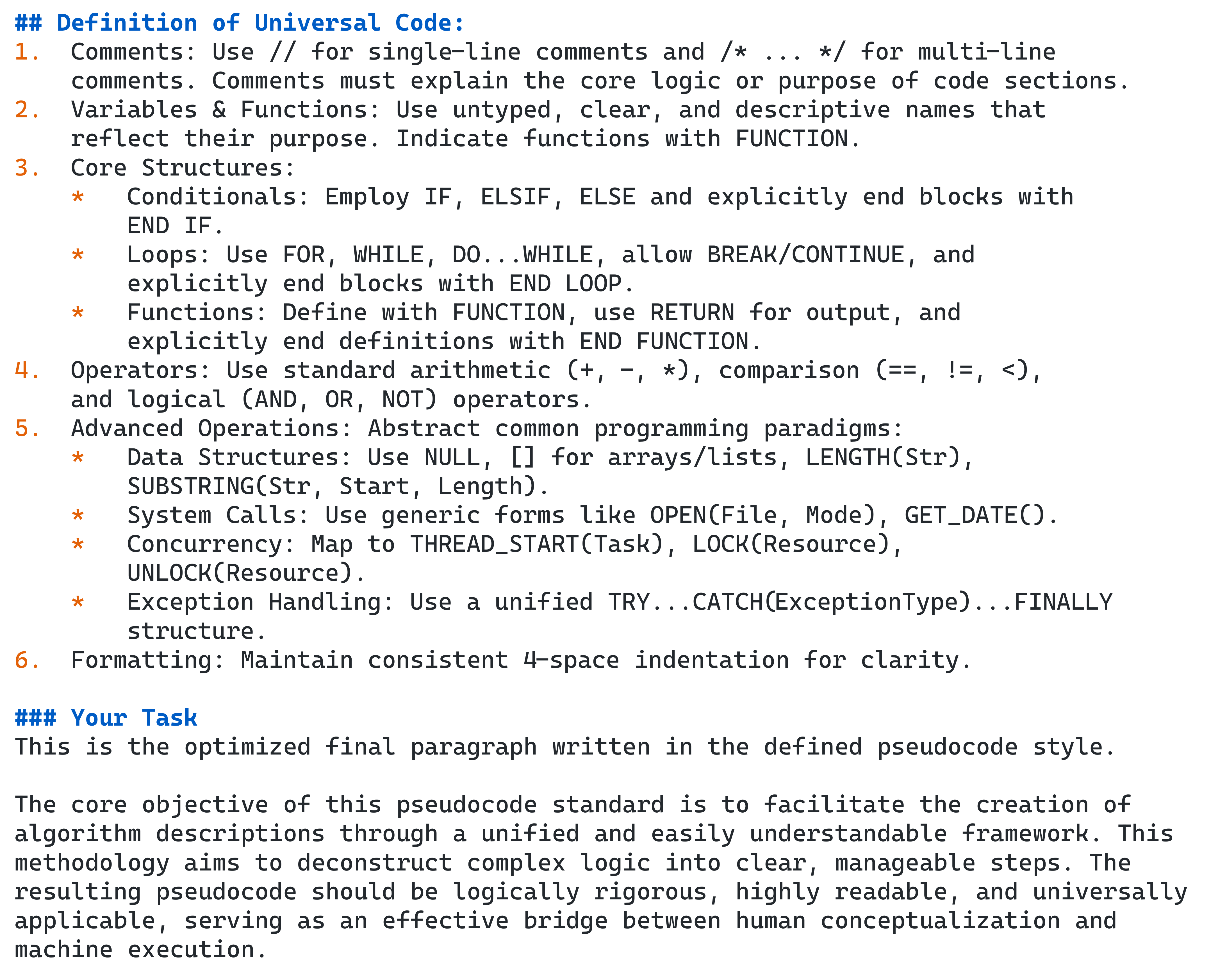} 
\caption{The prompt used to guide the LLM in generating our Universal Code, containing detailed specifications and an example.}
\label{fig:prompt}
\end{figure}

Figure~\ref{fig:prompt} details the prompt we used to guide the Large Language Model (LLM) in generating pseudocode, which includes the detailed specifications we have formulated.

The design advantage of this comprehensive specification lies in its \textit{Dual-Focus} capability. First, by mandating descriptive comments, variable names, and function names, it preserves the \textit{High-Level Semantic Intent} of the code. This ensures that the core logic and design philosophy ("what it does" and "why it does it") are not lost during abstraction, directly addressing the issue of critical textual information loss often seen in AST or IR representations due to oversimplification. Second, the specification introduces a set of strict, structured syntax, particularly explicit block terminators (e.g., \texttt{END IF}, \texttt{END LOOP}, \texttt{END FUNCTION}), to build an unambiguous syntactic framework. This design is crucial as it forces the mapping of variously formed syntactic structures from different programming languages (e.g., Python's indentation, C++/Java's curly braces) onto a unique \textit{Semantic Signature}. This systematic alignment effectively eliminates \textit{Semantic Interference} caused by syntactic differences, providing a solid foundation for the model to learn universal structural knowledge across languages. Furthermore, by uniformly abstracting advanced programming paradigms such as concurrency, exception handling, and file I/O, our pseudocode can more comprehensively cover the complexity of real-world code, ensuring the completeness and generalization ability of the representation. In summary, this design, which balances semantic richness with structural consistency, makes the generated pseudocode an ideal intermediate representation that maximally promotes efficient and lossless transfer of cross-lingual knowledge.

\subsubsection{Construction Process of Instruction-Following Dataset}
\label{sec:dataset_construction}

We employ a Large Language Model (LLM)\footnote{Specifically, we use Qwen3-Coder-30B-A3B-Instruct: \url{https://huggingface.co/Qwen/Qwen3-Coder-30B-A3B-Instruct}.} to automate the construction of a dataset containing the Universal Code representation. The core process follows the paradigm of "Construction From Instruction Dataset" and consists of the following steps:

First, for any given programming language $L$, we utilize an existing high-quality code instruction dataset $D_{L_s}$. This dataset consists of pairs $(q_\alpha, a_\alpha) \in D_{L_s}$, where $q_\alpha$ is a natural language question and $a_\alpha$ is the corresponding source code answer.

Second, through carefully designed Prompt Engineering, we guide an LLM to generate the corresponding Universal Code representation $p_\alpha$ for each pair. As shown in Figure~\ref{fig:prompt} and described in our methodology, our designed prompt template includes three key slots: \{Definition of Universal Code\}, \{Question\}, and \{Answer\}, which are filled with the predefined pseudocode specification, the original natural language query $q_\alpha$, and the source code answer $a_\alpha$, respectively.

Finally, we integrate the original query, source code, and the generated pseudocode to construct a Universal Code instruction dataset $D_{L_{u\alpha}}$ containing triplets $(q_\alpha, a_\alpha, p_\alpha)$.

We extend this process to $K$ different programming languages $L_{all} = \{L_k\}_{k=1}^K$, thereby creating a large-scale, multilingual Universal Code instruction dataset $D^*_{u\alpha} = \{D_{L_k^{u\alpha}}\}_{k=1}^K$. This dataset serves as the core training data for subsequent Supervised Fine-Tuning (SFT) of the model. In this study, we selected open-source instruction datasets as our starting point to ensure the reproducibility and fairness of our experiments.

\subsubsection{Quality Verification of Universal Code}
\label{sec:quality_verification}
To ensure the high-fidelity semantic preservation of the generated pseudocode, we employed a rigorous dual-verification mechanism:

\textbf{Structural Verification via AST:} We utilized Abstract Syntax Trees (AST) as a strict structural filter. Pseudocode that contradicts the control flow structure of the original code (e.g., missing crucial loops or conditional branches) is automatically discarded. This ensures the structural soundness of the dataset.

\textbf{Semantic Verification via Human Evaluation:} To validate full semantic equivalence, we conducted a human evaluation on 100 randomly sampled code-pseudocode pairs. Three senior PhD students annotated the samples, achieving an inter-annotator agreement (Cohen’s Kappa) of 0.79, which indicates strong consensus. Experts rated the samples on a 1-5 scale. The evaluation yielded a Semantic Preservation Score of 4.83, a Consistency Score of 4.65, and a Faithfulness Score of 4.92. These high scores confirm that our rigorous filtering pipeline successfully avoids the vast majority of generation errors.

\textbf{Error Analysis:} Through our manual analysis, we identified three primary error types that occasionally occur during LLM generation: (1) \textit{Misinterpretation of syntactic sugar}, where the model fails to accurately abstract language-specific shortcuts; (2) \textit{Hallucinated API expansions}, where the model incorrectly predicts the underlying implementation details of a high-level API; and (3) \textit{Scope ambiguity in closures}, where the variable scope within nested functions is inaccurately represented.

\subsubsection{Unified Contrastive Learning Pretraining Loss (UP)}
To align the semantic spaces represented by the triplets $(q_\alpha, a_\alpha, p_\alpha)$—natural language query, source code, and pseudocode—we introduce a contrastive learning method. The core idea is to pull closer the vectors of the query, code, and pseudocode generated from the same instruction triplet (positive pairs) in the embedding space, while pushing apart the vectors of elements from different triplets (negative pairs).

Specifically, we construct three pairwise contrastive learning tasks, using the pseudocode $p_\alpha$ as a bridge connecting the natural language $q_\alpha$ and the source code $a_\alpha$. We use the InfoNCE loss function to optimize the model~\cite{relate:contra1}. For a mini-batch of size $N$, the loss functions are defined as follows:

\textbf{Natural Language-Code Alignment Loss ($L_{q \leftrightarrow a}$):} This loss aims to align the natural language query $q_\alpha$ with its corresponding source code $a_\alpha$.
\begin{equation}
L_{q \leftrightarrow a} = - \sum_{i=1}^{N} \log \frac{\exp(\text{sim}(q_i, a_i) / \tau)}{\sum_{j=1}^{N} \exp(\text{sim}(q_i, a_j) / \tau)}
\end{equation}

\textbf{Natural Language-Pseudocode Alignment Loss ($L_{q \leftrightarrow p}$):} This loss aligns the natural language query $q_\alpha$ with its corresponding pseudocode representation $p_\alpha$.
\begin{equation}
L_{q \leftrightarrow p} = - \sum_{i=1}^{N} \log \frac{\exp(\text{sim}(q_i, p_i) / \tau)}{\sum_{j=1}^{N} \exp(\text{sim}(q_i, p_j) / \tau)}
\end{equation}

\textbf{Pseudocode-Code Alignment Loss ($L_{p \leftrightarrow a}$):} This loss treats the pseudocode $p_\alpha$ as an intermediate representation and aligns its semantics with the final source code $a_\alpha$.
\begin{equation}
L_{p \leftrightarrow a} = - \sum_{i=1}^{N} \log \frac{\exp(\text{sim}(p_i, a_i) / \tau)}{\sum_{j=1}^{N} \exp(\text{sim}(p_i, a_j) / \tau)}
\end{equation}

In the above equations:
\begin{itemize}[leftmargin=*]
    \item $(q_i, a_i, p_i)$ represent the vector embeddings of the $i$-th triplet in the batch.
    \item $\text{sim}(\cdot, \cdot)$ denotes the cosine similarity function.
    \item $\tau$ is a temperature hyperparameter that adjusts the smoothness of the probability distribution~\cite{DBLP:conf/nips/ZhangS18}.
    \item The summation in the denominator iterates over all negative samples in the batch (when $j \neq i$) and one positive sample (when $j = i$).
\end{itemize}

Finally, the total contrastive learning loss for the model is the sum of the three loss terms:
\begin{equation}
L_{\text{UP}} = L_{q \leftrightarrow a} + L_{q \leftrightarrow p} + L_{p \leftrightarrow a}
\end{equation}

By minimizing this total loss, the model learns a unified representation space where semantically equivalent natural language, source code, and pseudocode are mapped to nearby locations~\cite{SimCSE,DBLP:conf/emnlp/KarpukhinOMLWEC20}.

\subsection{Transfer Learning Stage}
In the pretraining stage, the model learns universal algorithmic logic that transcends language barriers through the unified pseudocode representation. However, to achieve precise retrieval in real-world multilingual codebases, the model must also deeply understand the unique syntactic features, keyword usage, and API call paradigms of each programming language. To this end, we design a multitask joint learning transfer stage aimed at efficiently and specifically transferring the general knowledge acquired during pretraining to fine-grained, language-specific representations.

The core idea of this stage is to decompose code into \textit{Code Slices} with different semantic functions and design a series of specialized contrastive and classification tasks. This forces the model to focus on different components of the code, thereby learning more robust and refined multilingual code representations.

\begin{figure}[h!]
\centering
\includegraphics[width=.9\linewidth]{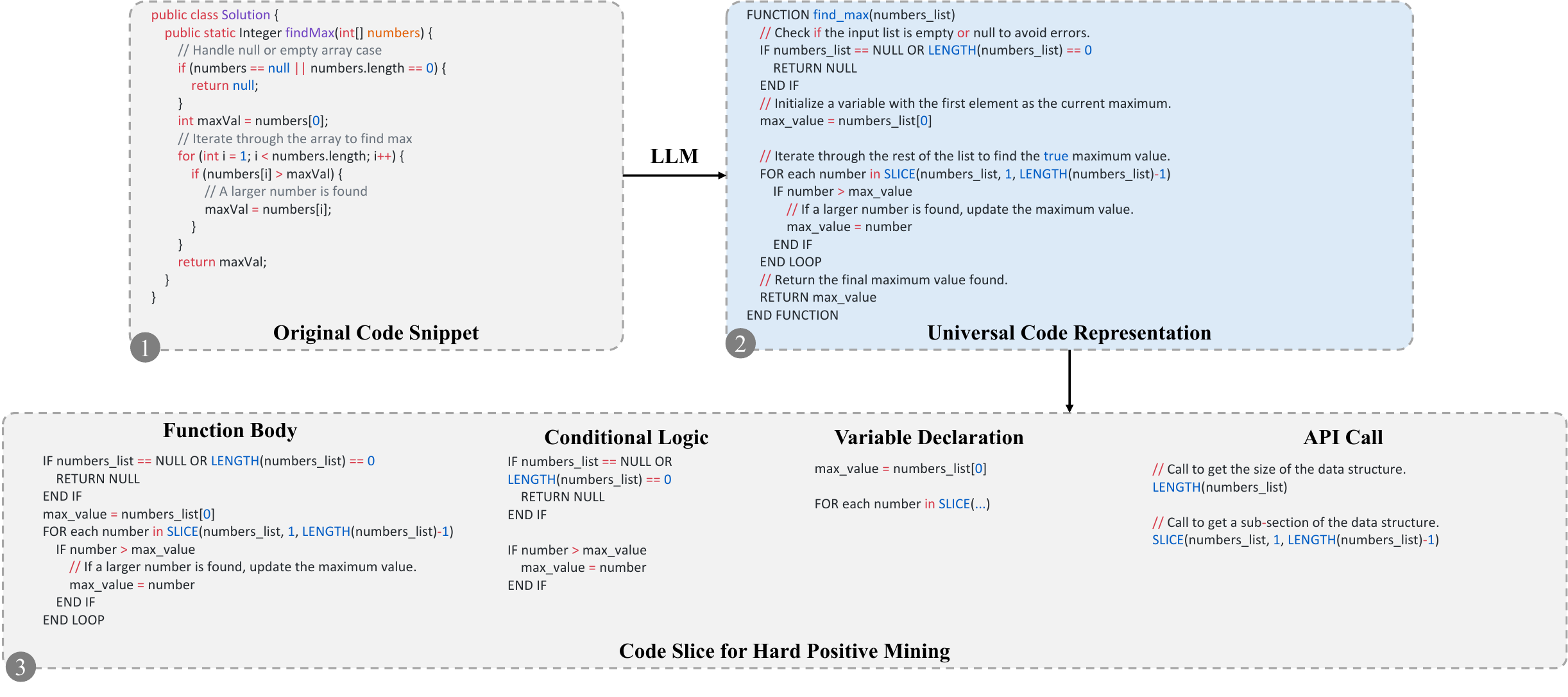} 
\caption{Examples of four semantic code slices.}
\label{fig:slice}
\end{figure}

\subsubsection{Semantic Code Slicing}
Traditional methods encode the entire function body as a single unit, which can lead the model to over-rely on surface-level features like function names, while neglecting the internal implementation logic, thereby harming generalization. To address this issue, we decompose source code functions into four types of slices, each carrying different semantic information, to ensure the model can understand the code from multiple dimensions. These four slice examples are shown in the Figure~\ref{fig:slice}

\begin{itemize}[leftmargin=*]
    \item \textbf{Function Body Slice:} We remove the function signature (including the function name and parameters), retaining only the function body. This forces the model to delve into the internal implementation logic of the code rather than relying on surface-level information from the interface.
    \item \textbf{Conditional Logic Slice:} We extract conditional branching statements like \texttt{if-else} and their corresponding code blocks. This helps the model capture the core execution paths and logic of the code.
    \item \textbf{Variable Declaration Slice:} We isolate the declaration and initialization parts of variables as a separate slice. This helps the model understand the data flow, lifecycle, and scope of variables in the code.
    \item \textbf{API Call Slice:} This part is crucial for multilingual code search. One of the most significant differences between programming languages is their distinct API ecosystems. For example, Python's \texttt{requests.get()} and JavaScript's \texttt{fetch()} are functionally equivalent but lexically completely different. Extracting API call sequences as a specialized slice for training forces the model to learn cross-lingual API functional alignment, breaking the semantic gap caused by different language ecosystems and enhancing its ability to decode high-density information containing numerous abbreviations and domain-specific terms.
\end{itemize}

\subsubsection{Multitask Joint Learning}
We designed a multitask joint learning framework that integrates hard positive mining, slice type detection, and dynamic hard negative mining. The original code, its four types of slices, and the natural language query are simultaneously fed into the model. Their output embedding vectors are jointly used to compute a combined loss function, optimizing the model collaboratively.

\paragraph{Hard Positive Mining Task (HP)}
This task aims to strengthen the model's understanding of "semantic invariance." We innovatively treat the natural language query $q$ and its corresponding four code slices $s_j$ (where $j \in \{\text{body, cond, var, api}\}$ ) as "hard positive" pairs. Since each slice contains only partial semantic information, the model, in order to align them with a query describing the full functionality, must learn more abstract and robust feature representations.

The loss function for this task, $L_{\text{HP}}$, is composed of multiple InfoNCE-based contrastive loss terms that use only in-batch negatives. For the $i$-th sample in a batch, its loss is:
\begin{equation}
L_{\text{HP}_i} = L_{\text{InfoNCE-IB}}(q_i, c_i) + \sum_{j \in S} L_{\text{InfoNCE-IB}}(q_i, s_{i,j})
\end{equation}
where $S = \{\text{body, cond, var, api}\}$, $c_i$ is the original complete code snippet, and $L_{\text{InfoNCE-IB}}$ is the InfoNCE loss using in-batch negatives:
\begin{equation}
L_{\text{InfoNCE-IB}}(x_i, y_i^+) = - \log \frac{\exp(\text{sim}(\mathbf{e}_{x_i}, \mathbf{e}_{y_i^+}) / \tau)}{
    \exp(\text{sim}(\mathbf{e}_{x_i}, \mathbf{e}_{y_i^+}) / \tau) + \sum_{y_j^- \in N_{ib}} \exp(\text{sim}(\mathbf{e}_{x_i}, \mathbf{e}_{y_j^-}) / \tau)
}
\end{equation}
Here, $N_{ib}$ represents the set of negative samples within the batch. The complete batch loss $L_{\text{HP}}$ is the average of all sample losses.

\paragraph{Code Slice Detection Task (SP)}
To enable the model to explicitly distinguish the semantic roles of different code components, we introduce an auxiliary classification task. A classification head is added on top of the model's output slice embeddings to predict their corresponding categories. This task is optimized using the standard Cross-Entropy Loss:
\begin{equation}
L_{\text{SP}} = - \frac{1}{B \cdot M} \sum_{i=1}^{B} \sum_{j=1}^{M} \sum_{k=1}^{K} y_{ij,k} \log(\hat{y}_{ij,k})
\end{equation}
where:
\begin{itemize}[leftmargin=*]
    \item $B$ is the batch size, $M=4$ is the number of slice types, and $K=4$ is the total number of classes.
    \item $y_{ij,k}$ is a one-hot vector indicating whether the $j$-th slice of the $i$-th sample belongs to class $k$.
    \item $\hat{y}_{ij,k}$ is the probability predicted by the model that it belongs to class $k$.
\end{itemize}

\paragraph{Hard Negative Mining Task (HN)}
To teach the model to distinguish between highly similar yet functionally different code snippets, we introduce a dedicated hard negative contrastive learning task. This is crucial for refining the model's decision boundaries in a dense multilingual embedding space~\cite{relate:contra2}.

We maintain a cross-lingual First-In-First-Out (FIFO) feature queue $Q$, which stores code and query embeddings from recent batches. For each positive pair $(q_i, c_i)$ in the current batch, we dynamically mine the top-$H$ most similar code snippets $\{c'_{i,k}\}_{k=1}^H$ from the queue to serve as hard negatives for the query $q_i$. Symmetrically, we also mine the top-$H$ most similar queries $\{q'_{i,k}\}_{k=1}^H$ as hard negatives for the code $c_i$~\cite{xia2021progcl,chu2021cuco}.

The hard negative contrastive loss, $L_{\text{HN}}$, is then defined as a sum of two symmetric terms. For each query $q_i$, we contrast its positive code pairing $c_i$ against its mined hard negative codes. The same is done for each code $c_i$ against its hard negative queries. The total loss for a batch of size $B$ is:
\begin{equation}
\begin{split}
L_{\text{HN}} = -\frac{1}{B} \sum_{i=1}^{B} \left( \log \frac{\exp(\text{sim}(q_i, c_i) / \tau)}{\exp(\text{sim}(q_i, c_i) / \tau) + \sum_{k=1}^{H} \exp(\text{sim}(q_i, c'_{i,k}) / \tau)} \right. \\
\left. + \log \frac{\exp(\text{sim}(c_i, q_i) / \tau)}{\exp(\text{sim}(c_i, q_i) / \tau) + \sum_{k=1}^{H} \exp(\text{sim}(c_i, q'_{i,k}) / \tau)} \right)
\end{split}
\end{equation}
This formulation directly forces the model to learn fine-grained distinctions by penalizing it for placing hard negatives too close to the query-code anchor pair~\cite{ding2020simplify,xie2023negative}.

\paragraph{Joint Learning}
Finally, we jointly optimize the aforementioned tasks. The total training loss $L_{\text{final}}$ is a weighted sum of the hard positive contrastive learning loss, the semantic slicing prediction loss, and the hard negative contrastive loss:
\begin{equation}
L_{\text{final}} = L_{\text{HP}} + \lambda_{SP} \cdot L_{\text{SP}} + \lambda_{HN} \cdot L_{\text{HN}}
\end{equation}
where $\lambda_{SP}$ and $\lambda_{HN}$ are hyperparameters that balance the importance of the three tasks. By minimizing $L_{\text{final}}$, the model not only inherits the general knowledge from the pretraining stage but also specifically learns the features, structures, and key differences of multilingual code, ultimately forming a powerful and balanced multilingual code retrieval engine.

\subsection{Code Search Stage (CS)}
After training is complete, we use the trained model for code search. We input a query $q$ and search for code that matches the intent within a given code repository $C = \{c_1, c_2, ..., c_n\}$.

Specifically, the trained model embeds the query $q$ and each code snippet $c_i$ in the repository $C$ into vectors $e_q$ and $e_{c_i}$, respectively. It then calculates the cosine similarity between $e_q$ and $e_{c_i}$ using the following formula:
\begin{equation}
\label{eq:cosine}
\text{CosineSimilarity}(q, c_i) = \frac{e_q \cdot e_{c_i}}{\|e_q\| \cdot \|e_{c_i}\|}
\end{equation}
Finally, the model ranks the code snippets based on their cosine similarity scores and outputs the top-$K$ results most relevant to the query $q$.

\section{Experimental Setup}
\label{sec:experimental_setup}

This section outlines the research questions guiding our evaluation, the datasets and baselines used, the evaluation metrics, and the experimental environment. Our goal is to assess the effectiveness and advancements of \projectName.

\subsection{Research Questions}
\begin{itemize}
    \item \textbf{RQ1: Multilingual Retrieval Capability.} How does \projectName perform in multilingual code search scenarios compared to state-of-the-art baselines?
    \item \textbf{RQ2: Cross-Lingual Retrieval Capability.} How effective is \projectName in cross-lingual search, where the query and code are in different languages?
    \item \textbf{RQ3: Transfer Learning to Niche Languages.} How well does \projectName perform on niche and low-resource programming languages in a zero-shot setting?
    \item \textbf{RQ4: Ablation Study.} What is the contribution of each key component of the \projectName framework?
\end{itemize}

\subsection{Datasets and Baselines}

\subsubsection{Datasets}

\paragraph{Mainstream Programming Languages} To evaluate retrieval capabilities in mainstream programming languages, we use CodeSearchNet, CoSQA, and APPS.
\textbf{CodeSearchNet(CSN)}~\cite{CodeSearchNet} contains six subdatasets for six programming languages (i.e., Ruby, Java, Python, JavaScript, Golang, PHP), where each data instance is a pair of a code snippet and its corresponding text description.
The \textbf{CoSQA}~\cite{huang-etal-2021-cosqa} dataset comprises 20,604 labeled pairs of natural language queries and codes, annotated by at least three human annotators.
\textbf{APPS}~\cite{hendrycks2021measuring} is a code generation benchmark with 10,000 problems, which we use to evaluate retrieval from natural language specifications.

\paragraph{Multilingual Code Retrieval} To evaluate multilingual code retrieval, we use data from XLCoST, Stack Overflow, and CodeFeedBack.
The \textbf{XLCoST}~\cite{zhu2022xlcost} dataset contains practical code search examples from GeeksForGeeks, covering eight languages. We utilize its natural language-to-code retrieval subset.
\textbf{CodeFeedBack-ST}~\cite{zheng2024opencodeinterpreter} consists of a corpus of 156k documents and 31k queries. \textbf{StackOverflow QA}~\cite{zheng2024opencodeinterpreter} reflects real-world developer queries, making it ideal for evaluating models on practical retrieval tasks.

\paragraph{Cross-lingual Code-to-Code Retrieval} To evaluate cross-lingual code-to-code retrieval, we use the \textbf{CodeTransOcean}~\cite{yan2023codetransocean} dataset. It is a large-scale benchmark supporting a wide variety of programming languages for code translation, including multilingual, niche, and deep learning framework translation tasks.

\paragraph{Niche Programming Languages} To evaluate the model's capabilities in niche programming languages, we collected a large dataset from GitHub~\cite{website:github}, covering 32 languages, including functional languages (Haskell, OCaml), older languages (Pascal, Fortran), and others. Following the CodeSearchNet strategy, we filtered out snippets with excessive whitespace, non-ASCII characters, and length < 3 lines to ensure high-quality semantic content. After filtering, 89k instances remain for testing our model's performance on these languages. More details are available in our repository.

\subsubsection{Baselines}
We compare \projectName with several powerful baseline models:
\begin{itemize}
    \item \textbf{openai-ada:}~\cite{neelakantan2022text} OpenAI’s highly efficient embedding model~\footnote{https://platform.openai.com/docs/guides/embeddings}.
    \item \textbf{text-embedding-3-small:}~\cite{neelakantan2022text} OpenAI's latest highly capable and cost-effective embedding model.
    \item \textbf{UniXCoder}~\cite{UniXcoder2022}: A unified cross-modal pre-trained model that leverages code, comments, and ASTs, showing strong performance in zero-shot code search.
    \item \textbf{Contriever}~\cite{izacard2021unsupervised}: A model from Facebook AI optimized for multilingual scenarios using in-batch negative sampling.
    \item \textbf{Code Retriever}~\cite{CodeRetriever2022}: Learns function-level code semantics through large-scale code-text contrastive pre-training.
    \item \textbf{BGE}~\cite{chen2024bge}: A state-of-the-art multilingual retrieval model designed for multi-linguality, multi-granularities, and multi-functionality.
\end{itemize}

\subsection{Metrics}
\label{subsec:metrics}
To comprehensively evaluate retrieval performance, we employ three widely adopted metrics: \textbf{Mean Reciprocal Rank (MRR)}, \textbf{Normalized Discounted Cumulative Gain (NDCG)}, and \textbf{Top@k} accuracy. MRR measures the average of the reciprocal ranks of the first relevant retrieved code snippet, reflecting the model's ability to return the correct answer at the very top. NDCG~\cite{busa2012apple} assesses the overall ranked list by assigning higher scores to more relevant items ranked higher, thus evaluating the general ranking quality. Finally, Top@k measures the percentage of queries where at least one correct code snippet appears in the top $k$ results, making it highly indicative of practical usability in developer-facing scenarios. We primarily report MRR, NDCG@10~\cite{NDCG}, and Top@10 across our benchmarks.

\subsection{Hyperparameters and Experimental Environment}
\label{subsec:hyperparameters}
We set the learning rate for the representation alignment stage to $1 \times 10^{-6}$ and for the multi-task joint learning stage to $2 \times 10^{-5}$. The temperature $\tau$ was set to 0.1, and the loss weighting coefficients $\lambda_{SP}$ and $\lambda_{HN}$ were both set to 0.5. We used the Adam optimizer with a batch size $B$ of 64. The queue size for dynamic hard negative mining was set to 3 times the batch size. We use UniXCoder-base~\cite{UniXcoder2022} as our base model. We employed an early stopping strategy, terminating training if performance on a validation set did not improve for 10 consecutive epochs, with a maximum of 100 epochs. All experiments were conducted on a machine equipped with eight 40G NVIDIA A100 GPUs.

\section{Experiments and Results}
\label{sec:experiments}

\subsection{RQ1: Multilingual Retrieval Capability}

\paragraph{Experimental Goal} To evaluate the overall performance of \projectName in multilingual code retrieval tasks, comparing its effectiveness and language balance against state-of-the-art baselines.

\paragraph{Experimental Design} We trained all models exclusively on the CodeSearchNet dataset. Evaluations were conducted in a zero-shot manner on seven diverse benchmarks: CodeSearchNet, CoSQA, APPS, StackOverflow QA, and CodeFeedBack (ST/MT). We used consistent hyperparameters and input lengths for all models. The primary metrics we examine include MRR, NDCG@k, and Top@10. Each experiment was repeated with three different random seeds, and we report the mean and standard deviation.

\paragraph{Experimental Results} As shown in Table~\ref{tab:multilingual_results}, \projectName demonstrates superior performance across all seven multilingual benchmarks. Evaluated comprehensively across robust multi-rank metrics (MRR, NDCG@10, and Top@10), \projectName achieves phenomenal combined average scores of 58.76\%, 54.44\%, and 67.93\% respectively. This constitutes a significant improvement of roughly 9.0 to 18.0 percentage points across metrics over the baselines. The performance gains are particularly pronounced on datasets with long-context queries and noisy real-world data, such as APPS, StackOverflow, and CodeFeedback, where the Top@10 accuracy consistently reaches excellent tiers (e.g., above 90\% on average). The results seamlessly validate the practicality and discriminative power of \projectName's semantic slicing and dynamic hard negative learning. Statistical significance tests confirm that these improvements are highly robust (p < 0.05).

\begin{mytitlebox}[Summary for RQ1]
\projectName achieves a new state-of-the-art in multilingual code retrieval. Its superior performance and language balance stem from the unified pseudo-code representation, which mitigates cross-lingual syntactic differences, and the semantic slicing and dynamic contrastive learning, which foster a deeper understanding of semantic invariance and clearer decision boundaries.
\end{mytitlebox}

\begin{table}[h!]
\centering
\caption{
Performance (MRR / NDCG@10 / Top@10, \%) on Multilingual Code Retrieval Benchmarks. All models were trained only on CodeSearchNet and evaluated in a zero-shot setting. \projectName consistently outperforms all baselines. The improvements of \projectName over the best baseline are statistically significant (p < 0.05). Note that MRR, NDCG, and Top@10 values are reported as percentages (\%).}
\label{tab:multilingual_results}
\resizebox{\textwidth}{!}{%
\begin{tabular}{llcccccccc}
\toprule
\textbf{Model} & \textbf{Metric} & \textbf{CSN} & \textbf{APPS} & \textbf{CoSQA} & \textbf{XLCoST(T2C)} & \textbf{SO QA} & \textbf{CF-ST} & \textbf{CF-MT} & \textbf{Average} \\
\midrule
\multirow{3}{*}{OpenAI-Ada-002}
& MRR    & 74.59 & 9.39  & 31.16 & 80.07 & 78.12 & 50.84 & 19.14 & 49.04 \\
& NDCG   & 69.13 & 8.70  & 28.88 & 74.21 & 72.40 & 47.12 & 17.74 & 45.45 \\
& Top@10 & 86.27 & 10.86 & 36.04 & 92.61 & 90.36 & 58.81 & 22.14 & 56.73 \\
\midrule
\multirow{3}{*}{text-embedding-3}
& MRR    & 76.32 & 9.95  & 32.85 & 81.15 & 79.68 & 52.11 & 19.62 & 50.24 \\
& NDCG   & 70.73 & 9.22  & 30.45 & 75.21 & 73.85 & 48.30 & 18.18 & 46.56 \\
& Top@10 & 88.27 & 11.51 & 38.00 & 93.86 & 92.17 & 60.28 & 22.69 & 58.11 \\
\midrule
\multirow{3}{*}{BGE-Base-en}
& MRR    & 49.16 & 4.37  & 35.35 & 75.10 & 79.36 & 70.12 & 33.90 & 49.62 \\
& NDCG   & 45.56 & 4.05  & 32.76 & 69.60 & 73.55 & 64.99 & 31.42 & 45.99 \\
& Top@10 & 56.86 & 5.05  & 40.88 & 86.86 & 91.79 & 81.11 & 39.21 & 57.39 \\
\midrule
\multirow{3}{*}{Contriever}
& MRR    & 38.56 & 5.55  & 15.33 & 37.46 & 71.27 & 59.46 & 42.33 & 38.56 \\
& NDCG   & 35.74 & 5.14  & 14.21 & 34.72 & 66.05 & 55.11 & 39.23 & 35.74 \\
& Top@10 & 44.60 & 6.41  & 17.73 & 43.33 & 82.43 & 68.78 & 48.96 & 44.60 \\
\midrule
\multirow{3}{*}{CodeRetriever}
& MRR    & 66.06 & 4.23  & 29.87 & 67.84 & 50.85 & 41.49 & 29.01 & 41.34 \\
& NDCG   & 61.22 & 3.92  & 27.68 & 62.87 & 47.13 & 38.45 & 26.89 & 38.31 \\
& Top@10 & 76.40 & 4.89  & 34.54 & 78.46 & 58.82 & 47.99 & 33.56 & 47.81 \\
\midrule
\multirow{3}{*}{UniXCoder}
& MRR    & 62.97 & 1.47  & 27.13 & 64.96 & 48.20 & 38.87 & 26.12 & 38.53 \\
& NDCG   & 58.36 & 1.36  & 25.14 & 60.20 & 44.67 & 36.02 & 24.21 & 35.71 \\
& Top@10 & 72.83 & 1.70  & 31.37 & 75.13 & 55.75 & 44.95 & 30.21 & 44.56 \\
\midrule
\multirow{3}{*}{\textbf{\projectName (Ours)}}
& MRR    & \textbf{78.20} & \textbf{10.58} & \textbf{36.50} & \textbf{84.79} & \textbf{81.94} & \textbf{73.80} & \textbf{45.53} & \textbf{58.76} \\
& NDCG   & \textbf{72.45} & \textbf{9.80}  & \textbf{33.82} & \textbf{78.56} & \textbf{75.92} & \textbf{68.37} & \textbf{42.18} & \textbf{54.44} \\
& Top@10 & \textbf{90.40} & \textbf{12.23} & \textbf{42.21} & \textbf{98.02} & \textbf{94.73} & \textbf{85.31} & \textbf{52.63} & \textbf{67.93} \\
\bottomrule
\end{tabular}
}
\end{table}

\subsection{RQ2: Cross-Lingual Retrieval Capability}

\paragraph{Experimental Goal} To assess the alignment and generalization capabilities of \projectName in cross-lingual retrieval scenarios, particularly on unseen language pairs and under real-world semantic divergences.

\paragraph{Experimental Design} We used the XLCoST (Text-to-Code subset, split by language) and CodeTransOcean datasets. Models were trained solely on CodeSearchNet's monolingual data and evaluated in a zero-shot cross-lingual setting. The retrieval index consisted of code in the target language, while queries were from a different source language (natural language or pseudocode). Performance was measured using NDCG@k.

\paragraph{Results and Discussion} Table~\ref{tab:crosslingual_results} shows that \projectName consistently leads in cross-lingual retrieval. On average, \projectName achieves an NDCG of 54.66, outperforming the strongest baselines by 3.49 to 10.85 points. The model shows consistent gains across various language pairs, including C++, Java, Python, and C\#. The advantage is particularly evident on the complex CodeTransOcean dataset, which features diverse programming frameworks and styles. This suggests that our approach—combining a unified pseudo-code representation with API call slicing—effectively mitigates the challenges posed by differing API ecosystems and naming conventions across languages. The unified representation bridges lexical gaps, while API slicing directly aligns semantically equivalent library calls. Furthermore, the dynamic cross-lingual hard negative mining strategy ensures that the model maintains sharp decision boundaries even among the most confusingly similar code snippets from different languages.

\begin{mytitlebox}[Summary for RQ2]
\projectName establishes a new state-of-the-art in zero-shot cross-lingual retrieval. Its success is attributed to a combination of three key factors: a unified pseudo-code representation that reduces lexical and stylistic noise, API call slicing that aligns functionally equivalent library calls, and a dynamic cross-lingual hard negative mining strategy that continuously refines the model's discriminative ability.
\end{mytitlebox}

\begin{table}[h!]
\centering
\caption{Performance (MRR / NDCG@10 / Top@10, \%) on Cross-Lingual Retrieval Benchmarks. \projectName shows robust performance in translating intent across language barriers. Note that metric values are reported as percentages (\%).}
\label{tab:crosslingual_results}
\resizebox{\textwidth}{!}{%
\begin{tabular}{llccccccccc}
\toprule
\textbf{Model} & \textbf{Metric} & \textbf{CodeTransOcean} & \textbf{XLCoST C++} & \textbf{XLCoST Java} & \textbf{XLCoST Py} & \textbf{XLCoST C\#} & \textbf{XLCoST JS} & \textbf{XLCoST PHP} & \textbf{XLCoST C} & \textbf{Average} \\
\midrule
\multirow{3}{*}{OpenAI-Ada-002}
& MRR    & 57.55 & 51.38 & 48.63 & 51.05 & 48.73 & 51.25 & 46.17 & 41.52 & 49.54 \\
& NDCG   & 53.34 & 47.62 & 45.07 & 47.31 & 45.16 & 47.50 & 42.79 & 38.48 & 45.91 \\
& Top@10 & 66.57 & 59.43 & 56.25 & 59.04 & 56.36 & 59.28 & 53.40 & 48.02 & 57.29 \\
\midrule
\multirow{3}{*}{BGE-Base-en-v1.5}
& MRR    & 41.54 & 52.78 & 51.61 & 52.99 & 52.05 & 53.50 & 48.20 & 43.29 & 49.50 \\
& NDCG   & 38.50 & 48.92 & 47.83 & 49.11 & 48.24 & 49.58 & 44.67 & 40.12 & 45.87 \\
& Top@10 & 48.05 & 61.05 & 59.69 & 61.29 & 60.20 & 61.88 & 55.75 & 50.07 & 57.25 \\
\midrule
\multirow{3}{*}{Contriever}
& MRR    & 47.65 & 47.82 & 47.67 & 48.78 & 49.99 & 49.76 & 45.05 & 41.47 & 47.27 \\
& NDCG   & 44.16 & 44.32 & 44.18 & 45.21 & 46.33 & 46.12 & 41.75 & 38.43 & 43.81 \\
& Top@10 & 55.11 & 55.31 & 55.14 & 56.42 & 57.82 & 57.56 & 52.10 & 47.96 & 54.68 \\
\midrule
\multirow{3}{*}{Code Retriever}
& MRR    & 66.06 & 55.41 & 54.89 & 55.48 & 55.34 & 55.22 & 51.61 & 47.71 & 55.22 \\
& NDCG   & 61.22 & 51.35 & 50.87 & 51.42 & 51.29 & 51.18 & 47.83 & 44.22 & 51.17 \\
& Top@10 & 76.40 & 64.08 & 63.49 & 64.17 & 64.01 & 63.87 & 59.69 & 55.19 & 63.86 \\
\midrule
\multirow{3}{*}{UniXCoder}
& MRR    & 45.12 & 52.56 & 52.76 & 51.75 & 52.46 & 52.40 & 47.93 & 44.54 & 49.94 \\
& NDCG   & 41.82 & 48.71 & 48.90 & 47.96 & 48.62 & 48.56 & 44.42 & 41.28 & 46.28 \\
& Top@10 & 52.19 & 60.79 & 61.03 & 59.85 & 60.68 & 60.60 & 55.44 & 51.52 & 57.76 \\
\midrule
\multirow{3}{*}{\textbf{\projectName (Ours)}}
& MRR    & \textbf{73.86}$^\ast$ & \textbf{61.31}$^\ast$ & \textbf{58.65}$^\ast$ & \textbf{58.18}$^\ast$ & \textbf{57.63}$^\ast$ & \textbf{58.56}$^\ast$ & \textbf{53.79}$^\ast$ & \textbf{49.83}$^\ast$ & \textbf{58.98}$^\ast$ \\
& NDCG   & \textbf{68.45}$^\ast$ & \textbf{56.82}$^\ast$ & \textbf{54.36}$^\ast$ & \textbf{53.92}$^\ast$ & \textbf{53.41}$^\ast$ & \textbf{54.27}$^\ast$ & \textbf{49.85}$^\ast$ & \textbf{46.18}$^\ast$ & \textbf{54.66}$^\ast$ \\
& Top@10 & \textbf{85.43}$^\ast$ & \textbf{70.91}$^\ast$ & \textbf{67.84}$^\ast$ & \textbf{67.29}$^\ast$ & \textbf{66.66}$^\ast$ & \textbf{67.73}$^\ast$ & \textbf{62.21}$^\ast$ & \textbf{57.63}$^\ast$ & \textbf{68.21}$^\ast$ \\
\bottomrule
\multicolumn{11}{l}{$^\ast$ \small{indicates improvement over the best baseline is statistically significant ($p < 0.05$) in the Approx. Randomization Test.}} \\
\end{tabular}
}
\end{table}

\subsection{RQ3: Transfer Learning to Niche Languages}

\paragraph{Experimental Goal} To evaluate the zero-shot retrieval capabilities of \projectName on niche and low-resource programming languages, verifying its ability to generalize robustly while maintaining performance balance.

\paragraph{Experimental Design} We used our self-curated \textbf{NicheLang} test set, which includes 32 low-resource languages such as Haskell, OCaml, Pascal, and Lua. Models were trained only on CodeSearchNet and evaluated on NicheLang in a zero-shot setting, without any language-specific fine-tuning.

\paragraph{Experimental Results} As shown in Table~\ref{tab:niche_lang_results}, \projectName demonstrates a remarkable ability to generalize to unseen, niche languages. Examining the comprehensive metrics, it achieves multi-metric average scores (MRR / NDCG@10 / Top@10) of 25.40\%, 23.53\%, and 29.36\%, significantly outperforming the next-best baseline, Code Retriever. The performance gains are completely unified across different language paradigms, including functional languages (Haskell, OCaml) and established procedural languages (Pascal, Fortran). We also incorporated state-of-the-art broad scale baseline \textit{text-embedding-3-small}, which improved representation stability, but still fundamentally lagged behind \projectName's specialized domain alignment. This confirms that our unified pseudo-code properly abstracts away syntactic idiosyncrasies, enabling the model to consistently rank equivalent algorithmic logics favorably. This generalization validates the effectiveness of our approach.

\begin{figure}[h!]
  \centering
  \includegraphics[width=.9\linewidth]{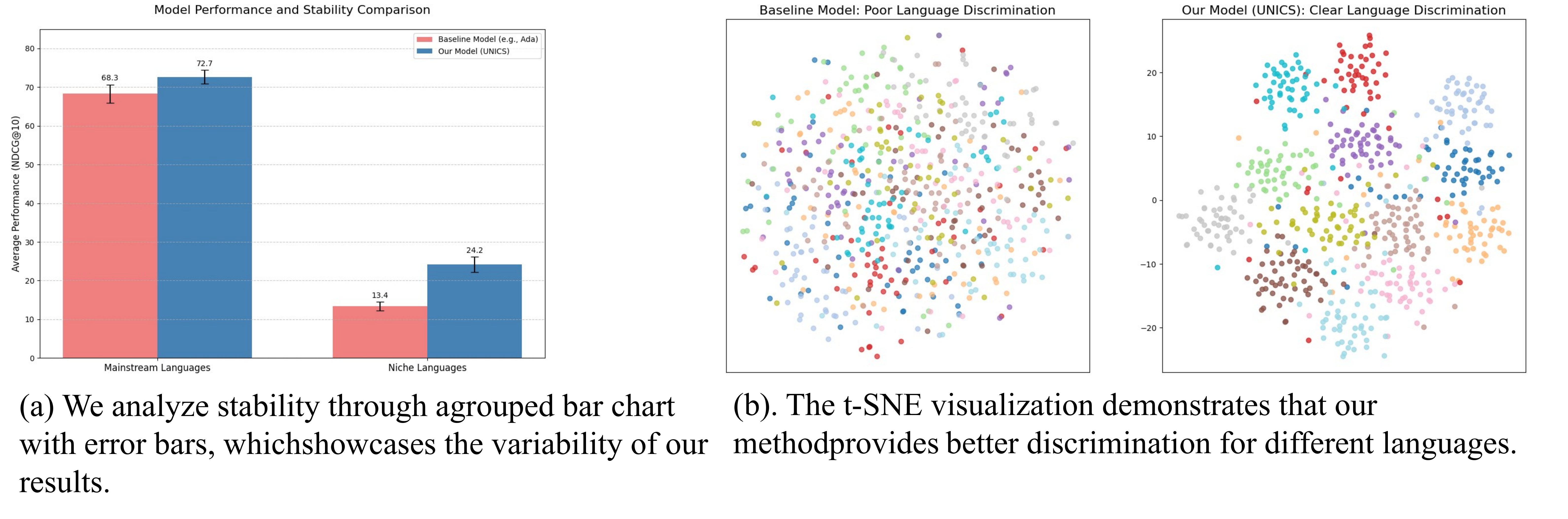} 
  \caption{Visualization of Performance and Language Discrimination in \projectName.}
  \label{fig:balance_matrix}
\end{figure}

\paragraph{Analysis of Stability and Language Discrimination}
As illustrated in our analysis, the model demonstrates both strong performance and cross-lingual discrimination. Figure~\ref{fig:balance_matrix} (left) presents a bar chart comparing the performance of \projectName against baselines on niche languages, visually reinforcing its superiority. Concurrently, the t-SNE visualization in Figure~\ref{fig:balance_matrix} (right) reveals that our approach yields well-separated clusters for different languages. This clear separation shows that our methodology successfully captures distinct linguistic features, creating a more robust and effective cross-lingual retrieval system.

\begin{mytitlebox}[Summary for RQ3]
\projectName achieves stable and significant zero-shot improvements on niche languages, demonstrating robust generalization to low-resource scenarios. This success is primarily due to the "de-lexicalized" unified pseudo-code, semantic slicing that highlights key programming mechanisms, and continuous discriminative learning from cross-lingual dynamic hard negatives.
\end{mytitlebox}

\begin{table}[h!]
\centering
\caption{Performance (MRR / NDCG@10 / Top@10, \%) on the Niche Programming Languages Benchmark (NicheLang). \projectName shows strong zero-shot generalization to unseen languages. Note that Metric values are reported as percentages (\%).}
\label{tab:niche_lang_results}
\resizebox{\textwidth}{!}{%
\begin{tabular}{llccccccccc}
\toprule
\textbf{Model} & \textbf{Metric} & \textbf{Haskell} & \textbf{OCaml} & \textbf{Scheme} & \textbf{Racket} & \textbf{Pascal} & \textbf{Fortran} & \textbf{Lua} & \textbf{Others} & \textbf{Average} \\
\midrule
\multirow{3}{*}{OpenAI-Ada-002}
& MRR    & 14.29 & 13.69 & 12.92 & 13.29 & 15.28 & 16.73 & 14.88 & 11.31 & 14.05 \\
& NDCG   & 13.24 & 12.68 & 11.97 & 12.31 & 14.16 & 15.50 & 13.79 & 10.48 & 13.02 \\
& Top@10 & 16.52 & 15.82 & 14.94 & 15.36 & 17.67 & 19.34 & 17.21 & 13.08 & 16.24 \\
\midrule
\multirow{3}{*}{text-embedding-3}
& MRR    & 15.10 & 14.45 & 13.58 & 14.01 & 16.10 & 17.63 & 15.68 & 11.92 & 14.81 \\
& NDCG   & 13.99 & 13.38 & 12.58 & 12.98 & 14.92 & 16.34 & 14.53 & 11.05 & 13.72 \\
& Top@10 & 17.46 & 16.70 & 15.70 & 16.19 & 18.62 & 20.38 & 18.14 & 13.79 & 17.12 \\
\midrule
\multirow{3}{*}{BGE-Base-en}
& MRR    & 15.65 & 15.02 & 16.01 & 15.23 & 17.53 & 18.98 & 16.91 & 12.00 & 15.92 \\
& NDCG   & 14.50 & 13.92 & 14.83 & 14.11 & 16.24 & 17.58 & 15.67 & 11.12 & 14.75 \\
& Top@10 & 18.09 & 17.37 & 18.50 & 17.61 & 20.26 & 21.94 & 19.55 & 13.88 & 18.40 \\
\midrule
\multirow{3}{*}{Contriever}
& MRR    & 9.89  & 8.98  & 7.75  & 8.86  & 11.15 & 12.00 & 10.52 & 6.94  & 9.51 \\
& NDCG   & 9.16  & 8.32  & 7.18  & 8.21  & 10.33 & 11.12 & 9.75  & 6.43  & 8.81 \\
& Top@10 & 11.43 & 10.38 & 8.96  & 10.24 & 12.89 & 13.88 & 12.17 & 8.02  & 11.00 \\
\midrule
\multirow{3}{*}{CodeRetriever}
& MRR    & 17.51 & 16.57 & 16.05 & 16.64 & 18.66 & 19.62 & 18.17 & 13.19 & 17.05 \\
& NDCG   & 16.22 & 15.35 & 14.87 & 15.42 & 17.29 & 18.18 & 16.83 & 12.22 & 15.80 \\
& Top@10 & 20.24 & 19.15 & 18.55 & 19.24 & 21.57 & 22.68 & 21.00 & 15.25 & 19.71 \\
\midrule
\multirow{3}{*}{UniXCoder}
& MRR    & 11.68 & 10.48 & 9.61  & 10.75 & 12.54 & 13.56 & 11.25 & 7.86  & 10.97 \\
& NDCG   & 10.82 & 9.71  & 8.90  & 9.96  & 11.62 & 12.56 & 10.42 & 7.28  & 10.16 \\
& Top@10 & 13.50 & 12.12 & 11.11 & 12.43 & 14.50 & 15.67 & 13.00 & 9.08  & 12.68 \\
\midrule
\multirow{3}{*}{\textbf{\projectName (Ours)}}
& MRR    & \textbf{25.31}$^\ast$ & \textbf{24.63}$^\ast$ & \textbf{23.05}$^\ast$ & \textbf{24.74}$^\ast$ & \textbf{28.51}$^\ast$ & \textbf{29.43}$^\ast$ & \textbf{26.82}$^\ast$ & \textbf{20.70}$^\ast$ & \textbf{25.40}$^\ast$ \\
& NDCG   & \textbf{23.45}$^\ast$ & \textbf{22.82}$^\ast$ & \textbf{21.36}$^\ast$ & \textbf{22.92}$^\ast$ & \textbf{26.41}$^\ast$ & \textbf{27.27}$^\ast$ & \textbf{24.85}$^\ast$ & \textbf{19.18}$^\ast$ & \textbf{23.53}$^\ast$ \\
& Top@10 & \textbf{29.26}$^\ast$ & \textbf{28.47}$^\ast$ & \textbf{26.65}$^\ast$ & \textbf{28.60}$^\ast$ & \textbf{32.95}$^\ast$ & \textbf{34.02}$^\ast$ & \textbf{31.00}$^\ast$ & \textbf{23.93}$^\ast$ & \textbf{29.36}$^\ast$ \\
\bottomrule
\end{tabular}
}
\end{table}

\subsection{RQ4: Model Ablation Study}

\paragraph{Experimental Goal} To validate the individual and cumulative contributions of each key component of \projectName (UP: Unified Pseudo-code; SP: Semantic Slicing; HP: Hard Positive contrastive learning; HN: cross-lingual dynamic Hard Negative mining) through an ablation study.

\paragraph{Experimental Design} We started with a base retrieval model and progressively added each \projectName component. We evaluated each variant on a representative subset of our benchmarks, including NicheLang, CodeTransOcean, XLCoST, CodeSearchNet, StackOverflow QA, and CodeFeedback-MT, using NDCG@k as the primary metric.

\paragraph{Experimental Results} The results of the ablation study, presented in Table~\ref{tab:ablation_results}, demonstrate the incremental benefits of each component.
Compared to the \textbf{Base Model}, we first evaluated a \textbf{Rule-based Pseudo-code (+ Rule-based UP)} generation method using AST parsing and regular expressions (e.g., snake\_case removal, explicit type abstraction). While providing some structural benefits, it achieved only 72.50\% MRR (corresponding to an estimated 67.19\% NDCG) on CodeSearchNet, structurally struggling to abstract complex logic into high-level intent. In contrast, replacing it with our LLM-driven \textbf{Unified Pseudo-code (+UP)} yields a much more significant and universal performance boost (average NDCG from 39.03 to 43.78, and normalizing API calls across languages). This confirms that the LLM-generated unified representation is critical for effectively reducing cross-lingual noise.
Introducing \textbf{Semantic Slicing (+SP)} further improves performance, particularly in tasks requiring nuanced understanding.
The addition of \textbf{Hard Positive mining (+HP)} brings another substantial gain (average NDCG to 51.06), especially on benchmarks with complex queries like StackOverflow and CodeFeedback, validating its role in learning semantic invariance.
Finally, the full \textbf{\projectName} model, which incorporates \textbf{dynamic Hard Negative mining (+HN)}, consistently achieves the highest performance. This final component solidifies the model's advantage, particularly on multilingual datasets like NicheLang and CodeTransOcean, by sharpening its discriminative capabilities against confusing cross-lingual examples and demonstrating the synergistic effect of all components.

\begin{mytitlebox}[Summary for RQ4]
Each component of \projectName provides a distinct and cumulative contribution. The unified pseudo-code and semantic slicing lay the foundation for cross-lingual semantic alignment. Hard positive mining enhances robust representation learning, while dynamic hard negative mining refines the model's ability to distinguish between closely related code snippets. Together, these components enable \projectName to achieve stable, state-of-the-art performance across a wide range of code retrieval tasks.
\end{mytitlebox}

\begin{table}[h!]
\centering
\caption{Ablation Study of \projectName Components (MRR / NDCG@10 / Top@10, \%). Each component provides a significant and cumulative performance improvement. (UP: Unified Pseudo-code Pretraining; SP: Semantic Slicing Prediction; HP: Hard Positive Contrastive Learning; HN: cross-lingual dynamic Hard Negative mining. * indicates estimated NDCG derived from the originally evaluated 72.50\% MRR score).}
\label{tab:ablation_results}
\resizebox{\textwidth}{!}{%
\begin{tabular}{llcccccccc}
\toprule
\textbf{Model Variant} & \textbf{Metric} & \textbf{NicheLang} & \textbf{CodeTransOcean} & \textbf{XLCoST(C2C)} & \textbf{CodeSearchNet} & \textbf{XLCoST(T2C)} & \textbf{SO QA} & \textbf{CodeFeedBack-MT} & \textbf{Average} \\
\midrule
\multirow{3}{*}{Base Model}
& MRR    & 13.11 & 43.43 & 50.60 & 62.00 & 46.17 & 52.31 & 27.20 & 42.12 \\
& NDCG   & 12.15 & 40.25 & 46.90 & 57.46 & 42.79 & 48.48 & 25.21 & 39.03 \\
& Top@10 & 15.16 & 50.23 & 58.53 & 71.71 & 53.40 & 60.50 & 31.46 & 48.71 \\
\midrule
\multirow{3}{*}{+ UP}
& MRR    & 17.05 & 47.13 & 53.20 & 67.70 & 52.83 & 57.76 & 35.04 & 47.24 \\
& NDCG   & 15.80 & 43.68 & 49.30 & 62.74 & 48.96 & 53.53 & 32.47 & 43.78 \\
& Top@10 & 19.72 & 54.51 & 61.53 & 78.30 & 61.10 & 66.81 & 40.52 & 54.64 \\
\midrule
\multirow{3}{*}{+ SP}
& MRR    & 18.07 & 48.71 & 54.11 & 69.01 & 55.43 & 59.47 & 36.77 & 48.80 \\
& NDCG   & 16.75 & 45.14 & 50.15 & 63.96 & 51.37 & 55.12 & 34.08 & 45.22 \\
& Top@10 & 20.90 & 56.33 & 62.59 & 79.82 & 64.11 & 68.79 & 42.53 & 56.44 \\
\midrule
\multirow{3}{*}{+ HP}
& MRR    & 22.60 & 53.65 & 55.51 & 73.25 & 70.32 & 68.68 & 41.67 & 55.10 \\
& NDCG   & 20.95 & 49.72 & 51.45 & 67.89 & 65.17 & 63.65 & 38.62 & 51.06 \\
& Top@10 & 26.15 & 62.05 & 64.21 & 84.73 & 81.33 & 79.44 & 48.20 & 63.73 \\
\midrule
\multirow{3}{*}{Full w/ RB-UP}
& MRR    & 24.63 & 67.38 & 52.54 & 72.50$^*$ & 78.29 & 74.36 & 41.20 & 58.70 \\
& NDCG   & 22.83 & 62.45 & 48.69 & 67.19$^*$ & 72.56 & 68.92 & 38.18 & 54.40 \\
& Top@10 & 28.49 & 77.94 & 60.76 & 83.85$^*$ & 90.55 & 86.01 & 47.65 & 67.89 \\
\midrule
\multirow{3}{*}{\textbf{\projectName (Full, with +HN)}}
& MRR    & \textbf{25.39} & \textbf{73.86} & \textbf{56.85} & \textbf{78.17} & \textbf{84.77} & \textbf{81.92} & \textbf{45.51} & \textbf{63.78} \\
& NDCG   & \textbf{23.53} & \textbf{68.45} & \textbf{52.69} & \textbf{72.45} & \textbf{78.56} & \textbf{75.92} & \textbf{42.18} & \textbf{59.11} \\
& Top@10 & \textbf{29.37} & \textbf{85.43} & \textbf{65.76} & \textbf{90.42} & \textbf{98.04} & \textbf{94.75} & \textbf{52.64} & \textbf{73.77} \\
\bottomrule
\end{tabular}
}
\end{table}

\subsection{Qualitative Error Analysis}
To conduct a qualitative analysis, we selected a representative set of samples from our test data. The selection process prioritized cases where \projectName and the baseline models exhibited divergent behavior. We focused on samples from niche programming languages (e.g., Rust, Lua, Swift) to highlight improvements in low-resource scenarios, while also including mainstream languages for comparison.

\paragraph{Success Case (Niche Language)} For a query in \textbf{Rust} asking "how to safely dereference a raw pointer," baseline models returned generic code snippets containing the `*` operator but ignored Rust's core "safety" constraint. In contrast, \projectName successfully retrieved a canonical code example using an `unsafe` block coupled with a null pointer check. This demonstrates that \projectName has a deeper understanding of language-specific philosophies and safety paradigms.

\paragraph{Failure Cases and Limitations} Despite its strong overall performance, our manual analysis of real-world retrieval errors by \projectName reveals three primary failure modes:
\begin{itemize}[leftmargin=*]
    \item \textbf{Hard Negatives:} The model occasionally fails to distinguish code snippets that are lexically very similar but functionally opposite (e.g., confusing $trim\_start$ vs. $trim\_end$).
    \item \textbf{API Mismatch:} When confronted with extremely rare or obscure third-party libraries, particularly in low-resource environments, the model sometimes fails to correctly link the specific API calls to the expected high-level algorithmic intent.
    \item \textbf{Test Code Interference:} In repository-level searches, test code or scripts often contain additional assertions, mock variables, and setup logistics. The model may misinterpret this supplementary information as core functionality, resulting in inaccurate retrieval.
\end{itemize}
These instances indicate that while the model excels at high-level semantic alignment, there remains room for improvement in handling fine-grained functional nuances, which clearly points out our future research directions.

\section{Threats to Validity}
\label{sec:threats}

Despite the demonstrated advantages of \projectName, our work is subject to several potential threats to validity that warrant consideration.

\paragraph{Construct Validity}
The accuracy of the generated pseudo-code cannot be fully guaranteed. Furthermore, it is uncertain whether this process introduces extraneous knowledge from the Large Language Model (LLM) used for generation.
However, we argue that our experimental design is fundamentally fair. Firstly, all pseudo-code undergoes verification via Abstract Syntax Tree (AST) slicing, which ensures its structural soundness. Secondly, even if some knowledge from the LLM is introduced, it is confined to fundamental programming paradigms. Therefore, we believe the improvements in multilingual capabilities are primarily attributable to our designed training tasks rather than the data itself.

\paragraph{Internal Validity}
We have not exhaustively explored the full space of hyperparameter configurations or alternative code slicing methods. Nevertheless, the current implementation has sufficiently demonstrated the effectiveness of our approach. It is plausible that a more comprehensive hyperparameter search could yield further performance gains.

\paragraph{External Validity}
While our proposed method is designed to be model-agnostic, we have not yet validated its efficacy on other model architectures (e.g., GPT-series models) or on models substantially larger than the 1B parameter scale. However, based on established trends in the field, we hypothesize that larger models would likely derive even more significant benefits from our approach.

\section{Conclusion and Future Work}
\label{sec:conclusion}

In this paper, we addressed the significant challenge of creating a unified code representation for effective multilingual and cross-lingual code retrieval. To this end, we introduced \projectName, a novel framework that leverages pseudo-code generation and a multi-task transfer learning strategy to align the semantic spaces of diverse programming languages. By employing a series of carefully designed pre-training tasks, including contrastive learning with dynamic hard negatives and hard positives, \projectName learns a robust unified representation that captures both high-level algorithmic logic and fine-grained structural details. Our extensive experiments on a wide range of benchmarks—spanning mainstream, multilingual, and niche programming languages—demonstrate that \projectName significantly outperforms existing state-of-the-art models, establishing a new benchmark for universal code embedding. Future work we will scale the \projectName framework to larger models (e.g., >10B parameters) and test its generalization across different architectures, including closed-source models like the GPT series. Additionally, we aim to continually expand our Niche Programming Language dataset to support an even broader spectrum of a developer's tooling.




\section{Data Availability}
All datasets and source code are publicly available~\cite{website:code_emb}. This repository includes the raw code, metadata, and filtering scripts for the NicheLang dataset.

\section*{Acknowledgments}
This work is supported by the National Science Foundation of China (92582204), and the 6th "333 Project" Leading Talent Team Project of Jiangsu Province. Jidong Ge is the corresponding author.

\bibliographystyle{ACM-Reference-Format}
\bibliography{main}

@inproceedings{DBLP:conf/emnlp/KarpukhinOMLWEC20,
  author       = {Vladimir Karpukhin and
                  Barlas Oguz and
                  Sewon Min and
                  Patrick S. H. Lewis and
                  Ledell Wu and
                  Sergey Edunov and
                  Danqi Chen and
                  Wen{-}tau Yih},
  editor       = {Bonnie Webber and
                  Trevor Cohn and
                  Yulan He and
                  Yang Liu},
  title        = {Dense Passage Retrieval for Open-Domain Question Answering},
  booktitle    = {Proceedings of the 2020 Conference on Empirical Methods in Natural
                  Language Processing, {EMNLP} 2020, Online, November 16-20, 2020},
  pages        = {6769--6781},
  publisher    = {Association for Computational Linguistics},
  year         = {2020},
  url          = {https://doi.org/10.18653/v1/2020.emnlp-main.550},
  doi          = {10.18653/V1/2020.EMNLP-MAIN.550},
  bibsource    = {dblp computer science bibliography, https://dblp.org}
}

@article{COCO2022old,
  author       = {Ensheng Shi and
                  Wenchao Gu and
                  Yanlin Wang and
                  Lun Du and
                  Hongyu Zhang and
                  Shi Han and
                  Dongmei Zhang and
                  Hongbin Sun},
  title        = {Enhancing Semantic Code Search with Multimodal Contrastive Learning
                  and Soft Data Augmentation},
  journal      = {CoRR},
  volume       = {abs/2204.03293},
  year         = {2022},
  url          = {https://doi.org/10.48550/arXiv.2204.03293},
  doi          = {10.48550/ARXIV.2204.03293},
  eprinttype    = {arXiv},
  eprint       = {2204.03293},
  bibsource    = {dblp computer science bibliography, https://dblp.org}
}

@inproceedings{COCO2022new,
    author = {Shi, Ensheng and Wang, Yanlin and Gu, Wenchao and Du, Lun and Zhang, Hongyu and Han, Shi and Zhang, Dongmei and Sun, Hongbin},
    title = {CoCoSoDa: Effective Contrastive Learning for Code Search},
    year = {2023},
    isbn = {9781665457019},
    publisher = {IEEE Press},
    url = {https://doi.org/10.1109/ICSE48619.2023.00185},
    doi = {10.1109/ICSE48619.2023.00185},
    booktitle = {Proceedings of the 45th International Conference on Software Engineering},
    pages = {2198–2210},
    numpages = {13},
    location = {Melbourne, Victoria, Australia},
    series = {ICSE '23}
}

@inproceedings{CodeBERT2020,
  author       = {Zhangyin Feng and
                  Daya Guo and
                  Duyu Tang and
                  Nan Duan and
                  Xiaocheng Feng and
                  Ming Gong and
                  Linjun Shou and
                  Bing Qin and
                  Ting Liu and
                  Daxin Jiang and
                  Ming Zhou},
  editor       = {Trevor Cohn and
                  Yulan He and
                  Yang Liu},
  title        = {CodeBERT: {A} Pre-Trained Model for Programming and Natural Languages},
  booktitle    = {Findings of the Association for Computational Linguistics: {EMNLP}
                  2020, Online Event, 16-20 November 2020},
  series       = {Findings of {ACL}},
  volume       = {{EMNLP} 2020},
  pages        = {1536--1547},
  publisher    = {Association for Computational Linguistics},
  year         = {2020},
  url          = {https://doi.org/10.18653/v1/2020.findings-emnlp.139},
  doi          = {10.18653/V1/2020.FINDINGS-EMNLP.139},
  bibsource    = {dblp computer science bibliography, https://dblp.org}
}

@inproceedings{CodeGPT2021,
  author       = {Shuai Lu and
                  Daya Guo and
                  Shuo Ren and
                  Junjie Huang and
                  Alexey Svyatkovskiy and
                  Ambrosio Blanco and
                  Colin B. Clement and
                  Dawn Drain and
                  Daxin Jiang and
                  Duyu Tang and
                  Ge Li and
                  Lidong Zhou and
                  Linjun Shou and
                  Long Zhou and
                  Michele Tufano and
                  Ming Gong and
                  Ming Zhou and
                  Nan Duan and
                  Neel Sundaresan and
                  Shao Kun Deng and
                  Shengyu Fu and
                  Shujie Liu},
  editor       = {Joaquin Vanschoren and
                  Sai{-}Kit Yeung},
  title        = {CodeXGLUE: {A} Machine Learning Benchmark Dataset for Code Understanding
                  and Generation},
  booktitle    = {Proceedings of the Neural Information Processing Systems Track on
                  Datasets and Benchmarks 1, NeurIPS Datasets and Benchmarks 2021, December
                  2021, virtual},
  year         = {2021},
  url          = {https://datasets-benchmarks-proceedings.neurips.cc/paper/2021/hash/c16a5320fa475530d9583c34fd356ef5-Abstract-round1.html},
  doi          = {10.48550/arXiv.2102.04664},
  bibsource    = {dblp computer science bibliography, https://dblp.org}
}

@inproceedings{CodeHow,
  author       = {Fei Lv and
                  Hongyu Zhang and
                  Jian{-}Guang Lou and
                  Shaowei Wang and
                  Dongmei Zhang and
                  Jianjun Zhao},
  editor       = {Myra B. Cohen and
                  Lars Grunske and
                  Michael Whalen},
  title        = {CodeHow: Effective Code Search Based on {API} Understanding and Extended
                  Boolean Model {(E)}},
  booktitle    = {30th {IEEE/ACM} International Conference on Automated Software Engineering,
                  {ASE} 2015, Lincoln, NE, USA, November 9-13, 2015},
  pages        = {260--270},
  publisher    = {{IEEE} Computer Society},
  year         = {2015},
  url          = {https://doi.org/10.1109/ASE.2015.42},
  doi          = {10.1109/ASE.2015.42},
  bibsource    = {dblp computer science bibliography, https://dblp.org}
}

@article{CodeRetriever2022,
  author       = {Xiaonan Li and
                  Yeyun Gong and
                  Yelong Shen and
                  Xipeng Qiu and
                  Hang Zhang and
                  Bolun Yao and
                  Weizhen Qi and
                  Daxin Jiang and
                  Weizhu Chen and
                  Nan Duan},
  title        = {CodeRetriever: Unimodal and Bimodal Contrastive Learning},
  journal      = {CoRR},
  volume       = {abs/2201.10866},
  year         = {2022},
  url          = {https://arxiv.org/abs/2201.10866},
  doi          = {10.48550/arXiv.2201.10866},
  eprinttype    = {arXiv},
  eprint       = {2201.10866},
  bibsource    = {dblp computer science bibliography, https://dblp.org}
}

@article{CodeSearchNet,
  author       = {Hamel Husain and
                  Ho{-}Hsiang Wu and
                  Tiferet Gazit and
                  Miltiadis Allamanis and
                  Marc Brockschmidt},
  title        = {CodeSearchNet Challenge: Evaluating the State of Semantic Code Search},
  journal      = {CoRR},
  volume       = {abs/1909.09436},
  year         = {2019},
  url          = {http://arxiv.org/abs/1909.09436},
  doi          = {10.48550/arXiv.1909.09436},
  eprinttype    = {arXiv},
  eprint       = {1909.09436},
  bibsource    = {dblp computer science bibliography, https://dblp.org}
}

@inproceedings{CodeT52021,
  author       = {Yue Wang and
                  Weishi Wang and
                  Shafiq R. Joty and
                  Steven C. H. Hoi},
  editor       = {Marie{-}Francine Moens and
                  Xuanjing Huang and
                  Lucia Specia and
                  Scott Wen{-}tau Yih},
  title        = {CodeT5: Identifier-aware Unified Pre-trained Encoder-Decoder Models
                  for Code Understanding and Generation},
  booktitle    = {Proceedings of the 2021 Conference on Empirical Methods in Natural
                  Language Processing, {EMNLP} 2021, Virtual Event / Punta Cana, Dominican
                  Republic, 7-11 November, 2021},
  pages        = {8696--8708},
  publisher    = {Association for Computational Linguistics},
  year         = {2021},
  url          = {https://doi.org/10.18653/v1/2021.emnlp-main.685},
  doi          = {10.18653/V1/2021.EMNLP-MAIN.685},
  bibsource    = {dblp computer science bibliography, https://dblp.org}
}

@inproceedings{ContraCode2021,
  author       = {Paras Jain and
                  Ajay Jain and
                  Tianjun Zhang and
                  Pieter Abbeel and
                  Joseph Gonzalez and
                  Ion Stoica},
  editor       = {Marie{-}Francine Moens and
                  Xuanjing Huang and
                  Lucia Specia and
                  Scott Wen{-}tau Yih},
  title        = {Contrastive Code Representation Learning},
  booktitle    = {Proceedings of the 2021 Conference on Empirical Methods in Natural
                  Language Processing, {EMNLP} 2021, Virtual Event / Punta Cana, Dominican
                  Republic, 7-11 November, 2021},
  pages        = {5954--5971},
  publisher    = {Association for Computational Linguistics},
  year         = {2021},
  url          = {https://doi.org/10.18653/v1/2021.emnlp-main.482},
  doi          = {10.18653/V1/2021.EMNLP-MAIN.482},
  bibsource    = {dblp computer science bibliography, https://dblp.org}
}

@inproceedings{CPM7,
  author       = {Alexey Svyatkovskiy and
                  Shao Kun Deng and
                  Shengyu Fu and
                  Neel Sundaresan},
  editor       = {Prem Devanbu and
                  Myra B. Cohen and
                  Thomas Zimmermann},
  title        = {IntelliCode compose: code generation using transformer},
  booktitle    = {{ESEC/FSE} '20: 28th {ACM} Joint European Software Engineering Conference
                  and Symposium on the Foundations of Software Engineering, Virtual
                  Event, USA, November 8-13, 2020},
  pages        = {1433--1443},
  publisher    = {{ACM}},
  year         = {2020},
  url          = {https://doi.org/10.1145/3368089.3417058},
  doi          = {10.1145/3368089.3417058},
  bibsource    = {dblp computer science bibliography, https://dblp.org}
}

@inproceedings{dahal2022scotch,
title={Scotch: A Semantic Code Search Engine for {IDE}s},
author={Samip Dahal and Adyasha Maharana and Mohit Bansal},
booktitle={Deep Learning for Code Workshop},
year={2022},
url={https://openreview.net/forum?id=rSxfCiOZk-c}
}

@inproceedings{DBLP:conf/iwpc/ShuaiX0Y0L20,
  author       = {Jianhang Shuai and
                  Ling Xu and
                  Chao Liu and
                  Meng Yan and
                  Xin Xia and
                  Yan Lei},
  title        = {Improving Code Search with Co-Attentive Representation Learning},
  booktitle    = {{ICPC} '20: 28th International Conference on Program Comprehension,
                  Seoul, Republic of Korea, July 13-15, 2020},
  pages        = {196--207},
  publisher    = {{ACM}},
  year         = {2020},
  url          = {https://doi.org/10.1145/3387904.3389269},
  doi          = {10.1145/3387904.3389269},
  bibsource    = {dblp computer science bibliography, https://dblp.org}
}

@inproceedings{DBLP:conf/kbse/WanSSXZ0Y19,
  author       = {Yao Wan and
                  Jingdong Shu and
                  Yulei Sui and
                  Guandong Xu and
                  Zhou Zhao and
                  Jian Wu and
                  Philip S. Yu},
  title        = {Multi-modal Attention Network Learning for Semantic Source Code Retrieval},
  booktitle    = {34th {IEEE/ACM} International Conference on Automated Software Engineering,
                  {ASE} 2019, San Diego, CA, USA, November 11-15, 2019},
  pages        = {13--25},
  publisher    = {{IEEE}},
  year         = {2019},
  url          = {https://doi.org/10.1109/ASE.2019.00012},
  doi          = {10.1109/ASE.2019.00012},
  bibsource    = {dblp computer science bibliography, https://dblp.org}
}

@inproceedings{DBLP:conf/nips/ZhangS18,
  author       = {Zhilu Zhang and
                  Mert R. Sabuncu},
  editor       = {Samy Bengio and
                  Hanna M. Wallach and
                  Hugo Larochelle and
                  Kristen Grauman and
                  Nicol{\`{o}} Cesa{-}Bianchi and
                  Roman Garnett},
  title        = {Generalized Cross Entropy Loss for Training Deep Neural Networks with
                  Noisy Labels},
  booktitle    = {Advances in Neural Information Processing Systems 31: Annual Conference
                  on Neural Information Processing Systems 2018, NeurIPS 2018, December
                  3-8, 2018, Montr{\'{e}}al, Canada},
  pages        = {8792--8802},
  year         = {2018},
  url          = {https://proceedings.neurips.cc/paper/2018/hash/f2925f97bc13ad2852a7a551802feea0-Abstract.html},
  doi          = {10.48550/arXiv.1805.07836},
  bibsource    = {dblp computer science bibliography, https://dblp.org}
}

@inproceedings{deepcs,
  author       = {Xiaodong Gu and
                  Hongyu Zhang and
                  Sunghun Kim},
  editor       = {Michel Chaudron and
                  Ivica Crnkovic and
                  Marsha Chechik and
                  Mark Harman},
  title        = {Deep code search},
  booktitle    = {Proceedings of the 40th International Conference on Software Engineering,
                  {ICSE} 2018, Gothenburg, Sweden, May 27 - June 03, 2018},
  pages        = {933--944},
  publisher    = {{ACM}},
  year         = {2018},
  url          = {https://doi.org/10.1145/3180155.3180167},
  doi          = {10.1145/3180155.3180167},
  bibsource    = {dblp computer science bibliography, https://dblp.org}
}

@inproceedings{GraphCodeBERT2021,
  author       = {Daya Guo and
                  Shuo Ren and
                  Shuai Lu and
                  Zhangyin Feng and
                  Duyu Tang and
                  Shujie Liu and
                  Long Zhou and
                  Nan Duan and
                  Alexey Svyatkovskiy and
                  Shengyu Fu and
                  Michele Tufano and
                  Shao Kun Deng and
                  Colin B. Clement and
                  Dawn Drain and
                  Neel Sundaresan and
                  Jian Yin and
                  Daxin Jiang and
                  Ming Zhou},
  title        = {GraphCodeBERT: Pre-training Code Representations with Data Flow},
  booktitle    = {9th International Conference on Learning Representations, {ICLR} 2021,
                  Virtual Event, Austria, May 3-7, 2021},
  publisher    = {OpenReview.net},
  year         = {2021},
  url          = {https://openreview.net/forum?id=jLoC4ez43PZ},
  doi          = {10.48550/arXiv.2009.08366},
  bibsource    = {dblp computer science bibliography, https://dblp.org}
}

@inproceedings{huang-etal-2021-cosqa,
  author       = {Junjie Huang and
                  Duyu Tang and
                  Linjun Shou and
                  Ming Gong and
                  Ke Xu and
                  Daxin Jiang and
                  Ming Zhou and
                  Nan Duan},
  editor       = {Chengqing Zong and
                  Fei Xia and
                  Wenjie Li and
                  Roberto Navigli},
  title        = {CoSQA: 20, 000+ Web Queries for Code Search and Question Answering},
  booktitle    = {Proceedings of the 59th Annual Meeting of the Association for Computational
                  Linguistics and the 11th International Joint Conference on Natural
                  Language Processing, {ACL/IJCNLP} 2021, (Volume 1: Long Papers), Virtual
                  Event, August 1-6, 2021},
  pages        = {5690--5700},
  publisher    = {Association for Computational Linguistics},
  year         = {2021},
  url          = {https://doi.org/10.18653/v1/2021.acl-long.442},
  doi          = {10.18653/V1/2021.ACL-LONG.442},
  bibsource    = {dblp computer science bibliography, https://dblp.org}
}

@inproceedings{ijcai2022p775,
  author       = {Changan Niu and
                  Chuanyi Li and
                  Bin Luo and
                  Vincent Ng},
  editor       = {Luc De Raedt},
  title        = {Deep Learning Meets Software Engineering: {A} Survey on Pre-Trained
                  Models of Source Code},
  booktitle    = {Proceedings of the Thirty-First International Joint Conference on
                  Artificial Intelligence, {IJCAI} 2022, Vienna, Austria, 23-29 July
                  2022},
  pages        = {5546--5555},
  publisher    = {ijcai.org},
  year         = {2022},
  url          = {https://doi.org/10.24963/ijcai.2022/775},
  doi          = {10.24963/IJCAI.2022/775},
  bibsource    = {dblp computer science bibliography, https://dblp.org}
}

@inproceedings{intro:deep1,
  author       = {Jos{\'{e}} Cambronero and
                  Hongyu Li and
                  Seohyun Kim and
                  Koushik Sen and
                  Satish Chandra},
  editor       = {Marlon Dumas and
                  Dietmar Pfahl and
                  Sven Apel and
                  Alessandra Russo},
  title        = {When deep learning met code search},
  booktitle    = {Proceedings of the {ACM} Joint Meeting on European Software Engineering
                  Conference and Symposium on the Foundations of Software Engineering,
                  {ESEC/SIGSOFT} {FSE} 2019, Tallinn, Estonia, August 26-30, 2019},
  pages        = {964--974},
  publisher    = {{ACM}},
  year         = {2019},
  url          = {https://doi.org/10.1145/3338906.3340458},
  doi          = {10.1145/3338906.3340458},
  bibsource    = {dblp computer science bibliography, https://dblp.org}
}

@inproceedings{intro:deep2,
  author       = {Saksham Sachdev and
                  Hongyu Li and
                  Sifei Luan and
                  Seohyun Kim and
                  Koushik Sen and
                  Satish Chandra},
  editor       = {Justin Gottschlich and
                  Alvin Cheung},
  title        = {Retrieval on source code: a neural code search},
  booktitle    = {Proceedings of the 2nd {ACM} {SIGPLAN} International Workshop on Machine
                  Learning and Programming Languages, MAPL@PLDI 2018, Philadelphia,
                  PA, USA, June 18-22, 2018},
  pages        = {31--41},
  publisher    = {{ACM}},
  year         = {2018},
  url          = {https://doi.org/10.1145/3211346.3211353},
  doi          = {10.1145/3211346.3211353},
  bibsource    = {dblp computer science bibliography, https://dblp.org}
}

@inproceedings{intro:deep3,
  author       = {Shuhan Yan and
                  Hang Yu and
                  Yuting Chen and
                  Beijun Shen and
                  Lingxiao Jiang},
  editor       = {Kostas Kontogiannis and
                  Foutse Khomh and
                  Alexander Chatzigeorgiou and
                  Marios{-}Eleftherios Fokaefs and
                  Minghui Zhou},
  title        = {Are the Code Snippets What We Are Searching for? {A} Benchmark and
                  an Empirical Study on Code Search with Natural-Language Queries},
  booktitle    = {27th {IEEE} International Conference on Software Analysis, Evolution
                  and Reengineering, {SANER} 2020, London, ON, Canada, February 18-21,
                  2020},
  pages        = {344--354},
  publisher    = {{IEEE}},
  year         = {2020},
  url          = {https://doi.org/10.1109/SANER48275.2020.9054840},
  doi          = {10.1109/SANER48275.2020.9054840},
  bibsource    = {dblp computer science bibliography, https://dblp.org}
}

@inproceedings{intro:deep4,
  author       = {Yitian Chai and
                  Hongyu Zhang and
                  Beijun Shen and
                  Xiaodong Gu},
  title        = {Cross-Domain Deep Code Search with Meta Learning},
  booktitle    = {44th {IEEE/ACM} 44th International Conference on Software Engineering,
                  {ICSE} 2022, Pittsburgh, PA, USA, May 25-27, 2022},
  pages        = {487--498},
  publisher    = {{ACM}},
  year         = {2022},
  url          = {https://doi.org/10.1145/3510003.3510125},
  doi          = {10.1145/3510003.3510125},
  bibsource    = {dblp computer science bibliography, https://dblp.org}
}

@article{intro:deep5,
  author       = {Shushan Arakelyan and
                  Anna Hakhverdyan and
                  Miltiadis Allamanis and
                  Luis Garcia and
                  Christophe Hauser and
                  Xiang Ren},
  title        = {{NS3:} Neuro-symbolic Semantic Code Search},
  booktitle    = {NeurIPS},
  year         = {2022},
  url          = {http://papers.nips.cc/paper\_files/paper/2022/hash/43f5f6c5cb333115914c8448b8506411-Abstract-Conference.html},
  doi          = {10.52202/068431-0761},
  bibsource    = {dblp computer science bibliography, https://dblp.org}
}

@inproceedings{intro:ir1,
  author       = {Joel Brandt and
                  Mira Dontcheva and
                  Marcos Weskamp and
                  Scott R. Klemmer},
  editor       = {Elizabeth D. Mynatt and
                  Don Schoner and
                  Geraldine Fitzpatrick and
                  Scott E. Hudson and
                  W. Keith Edwards and
                  Tom Rodden},
  title        = {Example-centric programming: integrating web search into the development
                  environment},
  booktitle    = {Proceedings of the 28th International Conference on Human Factors
                  in Computing Systems, {CHI} 2010, Atlanta, Georgia, USA, April 10-15,
                  2010},
  pages        = {513--522},
  publisher    = {{ACM}},
  year         = {2010},
  url          = {https://doi.org/10.1145/1753326.1753402},
  doi          = {10.1145/1753326.1753402},
  bibsource    = {dblp computer science bibliography, https://dblp.org}
}

@inproceedings{intro:ir10,
  author       = {Hongyu Zhang and
                  Anuj Jain and
                  Gaurav Khandelwal and
                  Chandrashekhar Kaushik and
                  Scott Ge and
                  Wenxiang Hu},
  editor       = {Thomas Zimmermann and
                  Jane Cleland{-}Huang and
                  Zhendong Su},
  title        = {Bing developer assistant: improving developer productivity by recommending
                  sample code},
  booktitle    = {Proceedings of the 24th {ACM} {SIGSOFT} International Symposium on
                  Foundations of Software Engineering, {FSE} 2016, Seattle, WA, USA,
                  November 13-18, 2016},
  pages        = {956--961},
  publisher    = {{ACM}},
  year         = {2016},
  url          = {https://doi.org/10.1145/2950290.2983955},
  doi          = {10.1145/2950290.2983955},
  bibsource    = {dblp computer science bibliography, https://dblp.org}
}

@inproceedings{intro:ir11,
  author       = {Jing Zhou and
                  Robert J. Walker},
  editor       = {Thomas Zimmermann and
                  Jane Cleland{-}Huang and
                  Zhendong Su},
  title        = {{API} deprecation: a retrospective analysis and detection method for
                  code examples on the web},
  booktitle    = {Proceedings of the 24th {ACM} {SIGSOFT} International Symposium on
                  Foundations of Software Engineering, {FSE} 2016, Seattle, WA, USA,
                  November 13-18, 2016},
  pages        = {266--277},
  publisher    = {{ACM}},
  year         = {2016},
  url          = {https://doi.org/10.1145/2950290.2950298},
  doi          = {10.1145/2950290.2950298},
  bibsource    = {dblp computer science bibliography, https://dblp.org}
}

@inproceedings{intro:ir12,
  author       = {Iman Keivanloo and
                  Juergen Rilling and
                  Ying Zou},
  editor       = {Pankaj Jalote and
                  Lionel C. Briand and
                  Andr{\'{e}} van der Hoek},
  title        = {Spotting working code examples},
  booktitle    = {36th International Conference on Software Engineering, {ICSE} '14,
                  Hyderabad, India - May 31 - June 07, 2014},
  pages        = {664--675},
  publisher    = {{ACM}},
  year         = {2014},
  url          = {https://doi.org/10.1145/2568225.2568292},
  doi          = {10.1145/2568225.2568292},
  bibsource    = {dblp computer science bibliography, https://dblp.org}
}

@inproceedings{intro:ir2,
  author       = {Brock Angus Campbell and
                  Christoph Treude},
  title        = {NLP2Code: Code Snippet Content Assist via Natural Language Tasks},
  booktitle    = {2017 {IEEE} International Conference on Software Maintenance and Evolution,
                  {ICSME} 2017, Shanghai, China, September 17-22, 2017},
  pages        = {628--632},
  publisher    = {{IEEE} Computer Society},
  year         = {2017},
  url          = {https://doi.org/10.1109/ICSME.2017.56},
  doi          = {10.1109/ICSME.2017.56},
  bibsource    = {dblp computer science bibliography, https://dblp.org}
}

@inproceedings{intro:ir3,
  author       = {Wing{-}Kwan Chan and
                  Hong Cheng and
                  David Lo},
  editor       = {Will Tracz and
                  Martin P. Robillard and
                  Tevfik Bultan},
  title        = {Searching connected {API} subgraph via text phrases},
  booktitle    = {20th {ACM} {SIGSOFT} Symposium on the Foundations of Software Engineering
                  (FSE-20), SIGSOFT/FSE'12, Cary, NC, {USA} - November 11 - 16, 2012},
  pages        = {10},
  publisher    = {{ACM}},
  year         = {2012},
  url          = {https://doi.org/10.1145/2393596.2393606},
  doi          = {10.1145/2393596.2393606},
  bibsource    = {dblp computer science bibliography, https://dblp.org}
}

@inproceedings{intro:ir4,
  author       = {Reid Holmes and
                  Rylan Cottrell and
                  Robert J. Walker and
                  J{\"{o}}rg Denzinger},
  title        = {The end-to-end use of source code examples: An exploratory study},
  booktitle    = {25th {IEEE} International Conference on Software Maintenance {(ICSM}
                  2009), September 20-26, 2009, Edmonton, Alberta, Canada},
  pages        = {555--558},
  publisher    = {{IEEE} Computer Society},
  year         = {2009},
  url          = {https://doi.org/10.1109/ICSM.2009.5306387},
  doi          = {10.1109/ICSM.2009.5306387},
  bibsource    = {dblp computer science bibliography, https://dblp.org}
}

@inproceedings{intro:ir5,
  author       = {Xuan Li and
                  Zerui Wang and
                  Qianxiang Wang and
                  Shoumeng Yan and
                  Tao Xie and
                  Hong Mei},
  editor       = {Thomas Zimmermann and
                  Jane Cleland{-}Huang and
                  Zhendong Su},
  title        = {Relationship-aware code search for JavaScript frameworks},
  booktitle    = {Proceedings of the 24th {ACM} {SIGSOFT} International Symposium on
                  Foundations of Software Engineering, {FSE} 2016, Seattle, WA, USA,
                  November 13-18, 2016},
  pages        = {690--701},
  publisher    = {{ACM}},
  year         = {2016},
  url          = {https://doi.org/10.1145/2950290.2950341},
  doi          = {10.1145/2950290.2950341},
  bibsource    = {dblp computer science bibliography, https://dblp.org}
}

@article{intro:ir7,
  author       = {Collin McMillan and
                  Mark Grechanik and
                  Denys Poshyvanyk and
                  Chen Fu and
                  Qing Xie},
  title        = {Exemplar: {A} Source Code Search Engine for Finding Highly Relevant
                  Applications},
  journal      = {{IEEE} Trans. Software Eng.},
  volume       = {38},
  number       = {5},
  pages        = {1069--1087},
  year         = {2012},
  url          = {https://doi.org/10.1109/TSE.2011.84},
  doi          = {10.1109/TSE.2011.84},
  bibsource    = {dblp computer science bibliography, https://dblp.org}
}

@inproceedings{intro:ir8,
  author       = {Collin McMillan and
                  Mark Grechanik and
                  Denys Poshyvanyk and
                  Qing Xie and
                  Chen Fu},
  editor       = {Richard N. Taylor and
                  Harald C. Gall and
                  Nenad Medvidovic},
  title        = {Portfolio: finding relevant functions and their usage},
  booktitle    = {Proceedings of the 33rd International Conference on Software Engineering,
                  {ICSE} 2011, Waikiki, Honolulu , HI, USA, May 21-28, 2011},
  pages        = {111--120},
  publisher    = {{ACM}},
  year         = {2011},
  url          = {https://doi.org/10.1145/1985793.1985809},
  doi          = {10.1145/1985793.1985809},
  bibsource    = {dblp computer science bibliography, https://dblp.org}
}

@inproceedings{intro:ir9,
  author       = {Luca Ponzanelli and
                  Gabriele Bavota and
                  Massimiliano Di Penta and
                  Rocco Oliveto and
                  Michele Lanza},
  editor       = {Premkumar T. Devanbu and
                  Sung Kim and
                  Martin Pinzger},
  title        = {Mining StackOverflow to turn the {IDE} into a self-confident programming
                  prompter},
  booktitle    = {11th Working Conference on Mining Software Repositories, {MSR} 2014,
                  Proceedings, May 31 - June 1, 2014, Hyderabad, India},
  pages        = {102--111},
  publisher    = {{ACM}},
  year         = {2014},
  url          = {https://doi.org/10.1145/2597073.2597077},
  doi          = {10.1145/2597073.2597077},
  bibsource    = {dblp computer science bibliography, https://dblp.org}
}

@article{intro:Linstead09Sourcer,
  author       = {Erik Linstead and
                  Sushil Krishna Bajracharya and
                  Trung Chi Ngo and
                  Paul Rigor and
                  Cristina Videira Lopes and
                  Pierre Baldi},
  title        = {Sourcerer: mining and searching internet-scale software repositories},
  journal      = {Data Min. Knowl. Discov.},
  volume       = {18},
  number       = {2},
  pages        = {300--336},
  year         = {2009},
  url          = {https://doi.org/10.1007/s10618-008-0118-x},
  doi          = {10.1007/S10618-008-0118-X},
  bibsource    = {dblp computer science bibliography, https://dblp.org}
}

@misc{website:github,
  title={Github Website},
  year={2023},
  howpublished="\url{https://www.github.com}"
}

@misc{website:code_emb,
  title={Code Embedding},
  year={2026},
  howpublished="\url{https://bitbucket.org/anonymous_code/code_embedding}"
}

@inproceedings{DBLP:conf/wcre/WangLM0J20,
  author       = {Wenhan Wang and
                  Ge Li and
                  Bo Ma and
                  Xin Xia and
                  Zhi Jin},
  editor       = {Kostas Kontogiannis and
                  Foutse Khomh and
                  Alexander Chatzigeorgiou and
                  Marios{-}Eleftherios Fokaefs and
                  Minghui Zhou},
  title        = {Detecting Code Clones with Graph Neural Network and Flow-Augmented
                  Abstract Syntax Tree},
  booktitle    = {27th {IEEE} International Conference on Software Analysis, Evolution
                  and Reengineering, {SANER} 2020, London, ON, Canada, February 18-21,
                  2020},
  pages        = {261--271},
  publisher    = {{IEEE}},
  year         = {2020},
  url          = {https://doi.org/10.1109/SANER48275.2020.9054857},
  doi          = {10.1109/SANER48275.2020.9054857},
  bibsource    = {dblp computer science bibliography, https://dblp.org}
}

@inproceedings{DBLP:conf/icsm/SvajlenkoIKRM14,
  author       = {Jeffrey Svajlenko and
                  Judith F. Islam and
                  Iman Keivanloo and
                  Chanchal Kumar Roy and
                  Mohammad Mamun Mia},
  title        = {Towards a Big Data Curated Benchmark of Inter-project Code Clones},
  booktitle    = {30th {IEEE} International Conference on Software Maintenance and Evolution,
                  Victoria, BC, Canada, September 29 - October 3, 2014},
  pages        = {476--480},
  publisher    = {{IEEE} Computer Society},
  year         = {2014},
  url          = {https://doi.org/10.1109/ICSME.2014.77},
  doi          = {10.1109/ICSME.2014.77},
  bibsource    = {dblp computer science bibliography, https://dblp.org}
}

@inproceedings{BrandtGLDK09,
  author       = {Joel Brandt and
                  Philip J. Guo and
                  Joel Lewenstein and
                  Mira Dontcheva and
                  Scott R. Klemmer},
  editor       = {Dan R. Olsen Jr. and
                  Richard B. Arthur and
                  Ken Hinckley and
                  Meredith Ringel Morris and
                  Scott E. Hudson and
                  Saul Greenberg},
  title        = {Two studies of opportunistic programming: interleaving web foraging,
                  learning, and writing code},
  booktitle    = {Proceedings of the 27th International Conference on Human Factors
                  in Computing Systems, {CHI} 2009, Boston, MA, USA, April 4-9, 2009},
  pages        = {1589--1598},
  publisher    = {{ACM}},
  year         = {2009},
  url          = {https://doi.org/10.1145/1518701.1518944},
  doi          = {10.1145/1518701.1518944},
  bibsource    = {dblp computer science bibliography, https://dblp.org}
}

@misc{zhu2022xlcost,
    title = {XLCoST: A Benchmark Dataset for Cross-lingual Code Intelligence},
    url = {https://arxiv.org/abs/2206.08474},
    doi = {10.48550/arXiv.2206.08474},
    author = {Zhu, Ming and Jain, Aneesh and Suresh, Karthik and Ravindran, Roshan and Tipirneni, Sindhu and Reddy, Chandan K.},
    year = {2022},
    eprint={2206.08474},
    archivePrefix={arXiv}
}

@article{neelakantan2022text,
  title={Text and code embeddings by contrastive pre-training},
  author={Neelakantan, Arvind and Xu, Tao and Puri, Raul and Radford, Alec and Han, Jesse Michael and Tworek, Jerry and Yuan, Qiming and Tezak, Nikolas and Kim, Jong Wook and Hallacy, Chris and others},
  journal={arXiv preprint arXiv:2201.10005},
  year={2022},
  doi={10.48550/arXiv.2201.10005}
}

@inproceedings{NDCG,
  author       = {Yining Wang and
                  Liwei Wang and
                  Yuanzhi Li and
                  Di He and
                  Tie{-}Yan Liu},
  editor       = {Shai Shalev{-}Shwartz and
                  Ingo Steinwart},
  title        = {A Theoretical Analysis of {NDCG} Type Ranking Measures},
  booktitle    = {{COLT} 2013 - The 26th Annual Conference on Learning Theory, June
                  12-14, 2013, Princeton University, NJ, {USA}},
  series       = {{JMLR} Workshop and Conference Proceedings},
  volume       = {30},
  pages        = {25--54},
  publisher    = {JMLR.org},
  year         = {2013},
  url          = {http://proceedings.mlr.press/v30/Wang13.html},
  doi          = {10.48550/arXiv.1304.6480},
  bibsource    = {dblp computer science bibliography, https://dblp.org}
}

@article{relate:contra1,
  author       = {Hongchao Fang and
                  Pengtao Xie},
  title        = {{CERT:} Contrastive Self-supervised Learning for Language Understanding},
  journal      = {CoRR},
  volume       = {abs/2005.12766},
  year         = {2020},
  url          = {https://arxiv.org/abs/2005.12766},
  doi          = {10.48550/arXiv.2005.12766},
  eprinttype    = {arXiv},
  eprint       = {2005.12766},
  bibsource    = {dblp computer science bibliography, https://dblp.org}
}

@inproceedings{relate:contra2,
  author       = {Nghi D. Q. Bui and
                  Yijun Yu and
                  Lingxiao Jiang},
  editor       = {Fernando Diaz and
                  Chirag Shah and
                  Torsten Suel and
                  Pablo Castells and
                  Rosie Jones and
                  Tetsuya Sakai},
  title        = {Self-Supervised Contrastive Learning for Code Retrieval and Summarization
                  via Semantic-Preserving Transformations},
  booktitle    = {{SIGIR} '21: The 44th International {ACM} {SIGIR} Conference on Research
                  and Development in Information Retrieval, Virtual Event, Canada, July
                  11-15, 2021},
  pages        = {511--521},
  publisher    = {{ACM}},
  year         = {2021},
  url          = {https://doi.org/10.1145/3404835.3462840},
  doi          = {10.1145/3404835.3462840},
  bibsource    = {dblp computer science bibliography, https://dblp.org}
}

@inproceedings{rerank21robust,
  author       = {Xiaofei Ma and
                  C{\'{\i}}cero Nogueira dos Santos and
                  Andrew O. Arnold},
  editor       = {Chengqing Zong and
                  Fei Xia and
                  Wenjie Li and
                  Roberto Navigli},
  title        = {Contrastive Fine-tuning Improves Robustness for Neural Rankers},
  booktitle    = {Findings of the Association for Computational Linguistics: {ACL/IJCNLP}
                  2021, Online Event, August 1-6, 2021},
  series       = {Findings of {ACL}},
  volume       = {{ACL/IJCNLP} 2021},
  pages        = {570--582},
  publisher    = {Association for Computational Linguistics},
  year         = {2021},
  url          = {https://doi.org/10.18653/v1/2021.findings-acl.51},
  doi          = {10.18653/V1/2021.FINDINGS-ACL.51},
  bibsource    = {dblp computer science bibliography, https://dblp.org}
}

@inproceedings{busa2012apple,
  title={An apple-to-apple comparison of learning-to-rank algorithms in terms of normalized discounted cumulative gain},
  author={Busa-Fekete, R{\'o}bert and Szarvas, Gy{\"o}rgy and Elteto, Tam{\'a}s and K{\'e}gl, Bal{\'a}zs},
  booktitle={ECAI 2012-20th European Conference on Artificial Intelligence: Preference Learning: Problems and Applications in AI Workshop},
  volume={242},
  year={2012},
  organization={Ios Press}
}

@article{chen2024bge,
  title={Bge m3-embedding: Multi-lingual, multi-functionality, multi-granularity text embeddings through self-knowledge distillation},
  author={Chen, Jianlv and Xiao, Shitao and Zhang, Peitian and Luo, Kun and Lian, Defu and Liu, Zheng},
  journal={arXiv preprint arXiv:2402.03216},
  year={2024},
  doi={10.48550/arXiv.2402.03216}
}

@article{yan2023codetransocean,
  title={Codetransocean: A comprehensive multilingual benchmark for code translation},
  author={Yan, Weixiang and Tian, Yuchen and Li, Yunzhe and Chen, Qian and Wang, Wen},
  journal={arXiv preprint arXiv:2310.04951},
  year={2023},
  doi={10.48550/arXiv.2310.04951}
}

@article{zheng2024opencodeinterpreter,
  title={Opencodeinterpreter: Integrating code generation with execution and refinement},
  author={Zheng, Tianyu and Zhang, Ge and Shen, Tianhao and Liu, Xueling and Lin, Bill Yuchen and Fu, Jie and Chen, Wenhu and Yue, Xiang},
  journal={arXiv preprint arXiv:2402.14658},
  year={2024},
  doi={10.48550/arXiv.2402.14658}
}

@article{hendrycks2021measuring,
  title={Measuring coding challenge competence with apps},
  author={Hendrycks, Dan and Basart, Steven and Kadavath, Saurav and Mazeika, Mantas and Arora, Akul and Guo, Ethan and Burns, Collin and Puranik, Samir and He, Horace and Song, Dawn and others},
  journal={arXiv preprint arXiv:2105.09938},
  year={2021},
  doi={10.48550/arXiv.2105.09938}
}

@article{izacard2021unsupervised,
  title={Unsupervised dense information retrieval with contrastive learning},
  author={Izacard, Gautier and Caron, Mathilde and Hosseini, Lucas and Riedel, Sebastian and Bojanowski, Piotr and Joulin, Armand and Grave, Edouard},
  journal={arXiv preprint arXiv:2112.09118},
  year={2021},
  doi={10.48550/arXiv.2112.09118}
}

@inproceedings{li2024consider,
  title={Consider: Commonalities and specialties driven multilingual code retrieval framework},
  author={Li, Rui and He, Liyang and Liu, Qi and Zhao, Yuze and Zhang, Zheng and Huang, Zhenya and Su, Yu and Wang, Shijin},
  booktitle={Proceedings of the AAAI Conference on Artificial Intelligence},
  volume={38},
  number={8},
  pages={8679--8687},
  year={2024},
  doi={10.1609/aaai.v38i8.28713}
}

@inproceedings{huang2023towards,
  title={Towards better multilingual code search through cross-lingual contrastive learning},
  author={Huang, Xiangbing and Ma, Yingwei and Zhou, Haifang and Jiang, Zhijie and Zhang, Yuanliang and Wang, Teng and Li, Shanshan},
  booktitle={Proceedings of the 14th Asia-Pacific Symposium on Internetware},
  pages={22--32},
  year={2023},
  doi={10.1145/3609437.3609439}
}

@inproceedings{ma2023mulcs,
  title={Mulcs: Towards a unified deep representation for multilingual code search},
  author={Ma, Yingwei and Yu, Yue and Li, Shanshan and Jia, Zhouyang and Ma, Jun and Xu, Rulin and Dong, Wei and Liao, Xiangke},
  booktitle={2023 IEEE International Conference on Software Analysis, Evolution and Reengineering (SANER)},
  pages={120--131},
  year={2023},
  organization={IEEE},
  doi={10.1109/SANER56733.2023.00021}
}

@inproceedings{SimCSE,
  author       = {Tianyu Gao and
                  Xingcheng Yao and
                  Danqi Chen},
  editor       = {Marie{-}Francine Moens and
                  Xuanjing Huang and
                  Lucia Specia and
                  Scott Wen{-}tau Yih},
  title        = {SimCSE: Simple Contrastive Learning of Sentence Embeddings},
  booktitle    = {Proceedings of the 2021 Conference on Empirical Methods in Natural
                  Language Processing, {EMNLP} 2021, Virtual Event / Punta Cana, Dominican
                  Republic, 7-11 November, 2021},
  pages        = {6894--6910},
  publisher    = {Association for Computational Linguistics},
  year         = {2021},
  url          = {https://doi.org/10.18653/v1/2021.emnlp-main.552},
  doi          = {10.18653/V1/2021.EMNLP-MAIN.552},
  bibsource    = {dblp computer science bibliography, https://dblp.org}
}

@article{SyncoBert2021,
 author = {Wang, Xin and Wang, Yasheng and Mi, Fei and Zhou, Pingyi and Wan, Yao and Liu, Xiao and Li, Li and Wu, Hao and Liu, Jin and Jiang, Xin},
 journal = {arXiv preprint arXiv:2108.04556},
 title = {Syncobert: Syntax-guided multi-modal contrastive pre-training for code representation},
 year = {2021},
 doi = {10.48550/arXiv.2108.04556}
}

@inproceedings{UniXcoder2022,
  author       = {Daya Guo and
                  Shuai Lu and
                  Nan Duan and
                  Yanlin Wang and
                  Ming Zhou and
                  Jian Yin},
  editor       = {Smaranda Muresan and
                  Preslav Nakov and
                  Aline Villavicencio},
  title        = {UniXcoder: Unified Cross-Modal Pre-training for Code Representation},
  booktitle    = {Proceedings of the 60th Annual Meeting of the Association for Computational
                  Linguistics (Volume 1: Long Papers), {ACL} 2022, Dublin, Ireland,
                  May 22-27, 2022},
  pages        = {7212--7225},
  publisher    = {Association for Computational Linguistics},
  year         = {2022},
  url          = {https://doi.org/10.18653/v1/2022.acl-long.499},
  doi          = {10.18653/V1/2022.ACL-LONG.499},
  bibsource    = {dblp computer science bibliography, https://dblp.org}
}

@inproceedings{xie2023negative,
  title={On negative sampling for contrastive audio-text retrieval},
  author={Xie, Huang and R{\"a}s{\"a}nen, Okko and Virtanen, Tuomas},
  booktitle={ICASSP 2023-2023 IEEE International Conference on Acoustics, Speech and Signal Processing (ICASSP)},
  pages={1--5},
  year={2023},
  organization={IEEE},
  doi={10.1109/ICASSP49357.2023.10095319}
}

@article{xia2021progcl,
  title={Progcl: Rethinking hard negative mining in graph contrastive learning},
  author={Xia, Jun and Wu, Lirong and Wang, Ge and Chen, Jintao and Li, Stan Z},
  journal={arXiv preprint arXiv:2110.02027},
  year={2021},
  doi={10.48550/arXiv.2110.02027}
}

@inproceedings{chu2021cuco,
  title={CuCo: Graph Representation with Curriculum Contrastive Learning.},
  author={Chu, Guanyi and Wang, Xiao and Shi, Chuan and Jiang, Xunqiang},
  booktitle={IJCAI},
  pages={2300--2306},
  year={2021},
  doi={10.24963/ijcai.2021/317}
}

@article{ding2020simplify,
  title={Simplify and robustify negative sampling for implicit collaborative filtering},
  author={Ding, Jingtao and Quan, Yuhan and Yao, Quanming and Li, Yong and Jin, Depeng},
  journal={Advances in Neural Information Processing Systems},
  volume={33},
  pages={1094--1105},
  year={2020},
  doi={10.48550/arXiv.2009.03376}
}

\end{document}